\documentclass[11pt]{article}
\pdfoutput=1
\usepackage{jheppub}

\usepackage{enumerate}
\usepackage{amsmath,amssymb,epsfig,cancel}
\usepackage{slashed}
\usepackage{color}
\usepackage{subfig}
\usepackage{bbold}
\usepackage{graphicx}
\usepackage{physics}
\usepackage{caption}
\usepackage{braket}
\usepackage[english]{babel}
\usepackage{pst-node,pst-dbicons}

\title{Spin-orbit duality}

\author{Kostas Filippas}

\affiliation{Institute of Nuclear and Particle Physics, NCSR Demokritos, GR-15310 Athens, Greece
}

\emailAdd{kfilippas21@gmail.com} 

\abstract{A new duality is proposed in four-dimensional flat space, which exchanges between spin and orbital degrees of freedom. This is motivated by a Hodge decomposition of the angular-momentum bivector for massive fields, along which spin and orbital angular momentum are Hodge duals of one another. The duality respects Poincar\`e symmetry and is shown to transform between complementary spacelike regions, projecting a fixed three-dimensional de Sitter world-tube (around the center of mass) into the bulk of four-dimensional spacetime and vice versa. This state of affairs is interpreted as a realization of the holographic principle. The dual theory living on that tube turns out to be noncommutative and entirely defined by the Casimir elements of the Poincar\`e algebra. In fact, the mass is now an ultraviolet cutoff. This naturally suggests that, for a Poincar\`e or just Lorentz-invariant quantum theory with massive fields of nonzero spin, spacetime is quantized at the fundamental level.
\\[5pt]
 }
 
\subheader{}
\vspace{6cm}
\dedicated{Dedicated to the memory of the Roma,\\Kostas Frangoulis and Nikos Sampanis.}

\begin{document}
\def\Tr{{\textrm{Tr}}}




\maketitle


\section{Introduction}
In this article, we present a duality in four-dimensional flat space between spin and orbital angular momentum. Those combine into a common conservation law $-$that of total angular momentum$-$ but, otherwise, are usually conceived to be different in nature. In other words, this is a map between internal and external degrees of freedom w.r.t. the spacetime manifold.\\

The motivation to consider such a bold connection springs from a stunning symmetry in the algebraic structure of angular momentum. The story unfolds as follows. First, Lorentz symmetry distinguishes between spacetime-independent (spin) and spacetime-dependent (orbital) angular momentum; this is just Noether's first theorem. The former is defined as the exterior algebra of position and momentum and is, hence, of kinematic nature. The latter is dynamic, in the sense that it is not kinematic and may interact with external spacelike fields. Those angular-momentum currents are complementary, since they both conspire to sustain a common conservation law. Following this kinematic-dynamic complementarity through, we project angular momentum on four-momentum. The logic is that kinematics may as well be defined as the projection along a kinematic four-vector, while dynamics should be whatever remains from that projection. This is achieved through the $(1+3)$ decomposition of tensors w.r.t. four-momentum. In turn, this yields a breakdown into what we call an electric and a magnetic part, titles which refer to the electromagnetic tensor whose usual decomposition into an electric and a magnetic field is actually tantamount to such a $(1+3)$ decomposition. We prove that, in the case of the total-angular-momentum tensor, orbital angular momentum is the electric and spin is the magnetic part. This decomposition reflects a structure that is manifestly symmetric under the exchange between those parts. Given this structural symmetry and motivated by Maxwell's electromagnetic duality, we show that this exchange between spin and orbit degrees of freedom respects Poincar\`e symmetry $-$in terms of the associated conservation laws and algebra$-$ and reflects a novel duality for every massive Poincar\`e or just Lorentz-invariant field theory. We call this the spin-orbit duality.\\

The implications of this duality are remarkable. For one, since orbital angular momentum is the exterior algebra on position and momentum, the duality implies an underlying map between four-position and the Pauli-Lubanski pseudovector. This transformation naturally selects a surface in spacetime $-$a three-dimensional de Sitter world-tube$-$ on which the duality becomes the trivial map. Surprisingly, the radius of this tube is identified with the so-called M\o{}ller radius found in the literature. This radius signifies a region of non-covariance inside of which only (pseudo-)worldlines of the center of mass live, while, quantum-mechanically, its eigenvalues are of the order of the Compton wavelength which poses a limitation on localization too. Hence, the duality map becomes trivial exactly where the center-of-mass position seizes to be valid anyway. At the same time, this world-tube is shown to be where the dual theory lives on. Thus, the duality not only interchanges spin with spatial degrees of freedom but also connects complementary regions of spacetime. This state of affairs, i.e. that spacelike information in four-dimensional bulk spacetime is encoded onto a three-dimensional de Sitter surface, eventually drives us to interpret this map as a holographic duality.\\

Yet, another striking feature here is that four-position transforms in such a way so that the dual quantum theory is a particular noncommutative space, defined entirely by the underlying Poincar\`e algebra. In fact, the dual noncommutative space may be classified as either the so-called spin-noncommutativity or a three-dimensional fuzzy de Sitter space. All of those algebras are already constructed in the literature. In the rest fame, its spatial subalgebra yields a fuzzy sphere, whereas we illustrate that orbital-angular-momentum `uncertainty rings' (around an axis) on this dual sphere correspond to spin (along the same axis) of quantum states in the bulk theory. It is those dual rings, accompanied by the realization of the duality as a hologram, which offer a good understanding for the exchange between internal (spin) and external (orbit) degrees of freedom across the dual pictures of field theory.\\

All this indicate that this duality may as well be seen as a link between large and small scales, i.e. the bulk spacetime and a de Sitter world-tube of the Compton scale. Therefore, since this world-tube is shown to be noncommutative, this naturally suggests that, for a Poincar\`e or just Lorentz-invariant quantum theory with massive fields of nonzero spin, spacetime is quantized at the fundamental level.\\

The plan of the article is the following. In Section \ref{SectionSpin} we review the Lorentz and Poincar\`e Noether currents and assign them proper interpretations. In Section \ref{SectionDuality} we prove the spin-orbit duality and extract its basic transformation properties. In Section \ref{SubsectionFuzzy} we show that the dual theory is noncommutative and present its fundamental features, while in Section \ref{SubsectionHoloInter} we give the duality map the interpretation of a hologram. Finally, we give out the reasons we think this duality has far-reaching consequences in Section \ref{SectionSummary}, while, in the end of the article, there are five Appendices with proofs of important statements made in the main text.

\section{Spin from Lorentz invariance}\label{SectionSpin}
Spin angular momentum is usually thought of as a quantum-mechanical property of particles. This is most often due to Wigner's classification of unitary irreducible representations of the Poincar\'e group \cite{Wigner:1939cj}, which are labeled by two parameters $m\geq0$ and $s=0,\frac{1}{2},1...$, respectively realized as the mass and spin of representations. However, as e.g. mass or orbital angular momentum are present in both the classical and quantum level, so is spin; discretization in the quantum regime does not preclude a classical analogy.

\subsection{Conservation law}
In the most basic level, spin angular momentum silently emerges just from the pure transformation properties of the Lorentz algebra, which acts on fields through the generator $\mathbf{M}_{\mu\nu}=\mathbf{S}_{\mu\nu}+\mathbf{L}_{\mu\nu}$. The familiar differential operator $\mathbf{L}_{\mu\nu}\equiv -i(x_{\mu}\,\partial_{\nu}-x_{\nu}\,\partial_{\mu})$ is an orbital rotation, while the four-vector representation ${(\mathbf{S}^{\mu\nu})^\rho}_\sigma\equiv -i(g^{\mu\rho}{\delta^\nu}_\sigma-g^{\nu\rho}{\delta^\mu}_\sigma)$, or the spinor representation ${(\mathbf{S}^{\mu\nu})}_{\alpha\beta}=-\frac{i}{2}[\gamma^\mu,\gamma^\nu]_{\alpha\beta}$, reflect how the field changes when frames change; $\mathbf{S}_{\mu\nu}=0$ is understood for scalar fields. In order to prove rigorously that $\mathbf{S}_{\mu\nu}$ is associated with spin and $\mathbf{M}_{\mu\nu}$ with the total angular momentum, we consider those transformations as symmetries. In fact, we assume invariance under the larger ISO$(1,3)$ Poincar\'e group, where, given a generic action principle of the form $S=S[q,\partial q]=\int\dd^4x\,\mathcal{L}[q,\partial q]$, symmetry manifests through Noether's archetypal first theorem. When the equations of motion are satisfied, the translations subgroup implies the conserved current density

\begin{equation}
\mathcal{T}^{\mu\nu}\;\equiv\;\frac{\partial\mathcal{L}}{\partial\partial_\mu q}\partial^\nu q-g^{\mu\nu}\mathcal{L}\;, \hspace{1cm}\partial_\mu \mathcal{T}^{\mu\nu}=0\hspace{1cm}\Rightarrow\hspace{1cm}\dot{p}^\mu\;=\;\int\dd^3x\,\mathcal{T}^{0\mu}\;=\;0\;,\label{CanEMTensor}
\end{equation}\\
where $\mathcal{T}^{\mu\nu}$ is the canonical energy-momentum tensor and $p^\mu$ the total four-momentum. This current, in general, is not symmetric. Next, Lorentz rotations imply the conserved current density
\begin{equation}
\mathcal{M}^{\rho\mu\nu}\;\equiv\;i\,\frac{\partial\mathcal{L}}{\partial\partial_\rho q}\,\mathbf{M}^{\mu\nu}q\;=\;i\,\frac{\partial\mathcal{L}}{\partial\partial_\rho q}\,\mathbf{S}^{\mu\nu}q\,+\,\left(x^\mu\,\mathcal{T}^{\rho\nu}-x^\nu\,\mathcal{T}^{\rho\mu}\right)\;,\hspace{1cm}\partial_\rho\,\mathcal{M}^{\rho\mu\nu}\;=\;0\;.
\end{equation}\\
This integrates over space to the total angular momentum of the system,

\begin{equation}
M^{\mu\nu}\;=\;\int\dd^3x\,\mathcal{M}^{0\mu\nu}\;=\;\int\dd^3x\,\left(i\,\frac{\partial\mathcal{L}}{\partial\dot{q}}\,\mathbf{S}^{\mu\nu}q\,+\,\left(x^\mu\,\mathcal{T}^{0\nu}-x^\nu\,\mathcal{T}^{0\mu}\right)\right)\;,\label{TotalAngMom}
\end{equation}\\
providing a natural separation between two kinds of angular-momentum degrees of freedom. Those are a spacetime-dependent and a spacetime-independent term,

\begin{equation}
\mathcal{L}^{\mu\nu}\;\equiv\;\mathcal{L}^{0\mu\nu}\;\equiv\;x^\mu\,\mathcal{T}^{0\nu}-x^\nu\,\mathcal{T}^{0\mu}\;,\hspace{2cm}\mathcal{S}^{\mu\nu}\;\equiv\;\mathcal{S}^{0\mu\nu}\;\equiv\;i\,\frac{\partial\mathcal{L}}{\partial\dot{q}}\,\mathbf{S}^{\mu\nu}q\;,\label{OrbitDens}
\end{equation}\\
which are what we call the \textit{orbital} and \textit{spin-angular-momentum} densities, respectively. Total current conservation, $\partial_\rho\mathcal{M}^{\rho\mu\nu}=0$, gives

\begin{equation}
\partial_\rho\,\mathcal{L}^{\rho\mu\nu}\;=\;-\partial_\rho\,\mathcal{S}^{\rho\mu\nu}\;=\;\mathcal{T}^{\mu\nu}-\mathcal{T}^{\nu\mu}\;,\label{TotalAngMomCons}
\end{equation}\\
which is just the statement that the presence of spin in a system implies an asymmetric canonical energy-momentum tensor. Accordingly, orbital angular momentum is not independently conserved but cancels out with spin angular momentum, their total quantities related as

\begin{equation}
\dot{M}^{\mu\nu}\;=\;\dot{L}^{\mu\nu}+\dot{S}^{\mu\nu}\;=\;0\;.\label{TotalAngMomCons2}
\end{equation}
This tension between the two kinds of angular momentum inside their conservation law makes us wonder whether spin could be realized too as a form of orbital angular momentum.

\subsection{The full energy-momentum current}\label{SectionEMtensor}
To answer this question, we should seek for an energy-momentum tensor $\Theta^{\mu\nu}$, such that all angular momentum in the system (i.e. including spin) will be deduced from its rotational flow. That is, in view of (\ref{TotalAngMom}), we want instead

\begin{equation}
M^{\mu\nu}\;\stackrel{!}{=}\;\int\dd^3x\,\left(x^\mu\,\Theta^{0\nu}-x^\nu\,\Theta^{0\mu}\right)\;,\label{BelTotalAngMom}
\end{equation}\\
which would yield the same total angular momentum as in (\ref{TotalAngMom}). This implies the also-improved current

\begin{equation}
\mathcal{M}_\Theta^{\rho\mu\nu}=x^\mu\Theta^{\rho\nu}-x^\nu\Theta^{\rho\mu}\;\;\neq\;\mathcal{M}^{\rho\mu\nu}\;,
\end{equation}\\
which, in turn, integrates over space to give the same total angular momentum $M_\Theta^{\mu\nu}=M^{\mu\nu}$. Then, conservation of the new current $\mathcal{M}_\Theta^{\rho\mu\nu}$,

\begin{equation}
\partial_\rho\mathcal{M}_\Theta^{\rho\mu\nu}\;=\;\Theta^{\mu\nu}-\Theta^{\nu\mu}\;=\;0\;,
\end{equation}\\
implies that we want a symmetric energy-momentum current. This symmetrization is achieved by employing the Lorentz symmetry to eliminate the six antisymmetric components of the canonical energy-momentum tensor. Such redefinition may seem arbitrary but is certainly not. This is because in the usual application of Noether's first theorem an integration by parts takes place and, thus, the canonical current is always ambiguous under addition of terms of the form $\partial_\rho U^{[\rho\mu]\nu}$. In fact, this procedure has been worked out by Belinfante \cite{Belinfante:1939} and Rosenfeld \cite{Rosenfeld:1940} and results in the unique current

\begin{equation}
\Theta^{\mu\nu}\;=\;\mathcal{T}^{\mu\nu}\,+\,\frac{1}{2}\partial_\rho\left(\mathcal{S}^{\rho\mu\nu}\,+\,\mathcal{S}^{\nu\mu\rho}\,+\,\mathcal{S}^{\mu\nu\rho}\right)\;.\label{BelinTensor}
\end{equation}\\
The new energy-momentum current $\Theta^{\mu\nu}$ is manifestly conserved, yields the same total energy-momentum and, as a bonus, is symmetric,

\begin{equation}
\partial_\mu\,\Theta^{\mu\nu}=0\;,\hspace{1cm}p^\nu=\int\dd^3x\,\Theta^{0\nu}=\int\dd^3x\,\mathcal{T}^{0\nu}\;,\hspace{1cm}\Theta^{\mu\nu}=\Theta^{\nu\mu}\;,
\end{equation}\\
where being symmetric is precisely the result of incorporating spin into the picture. In Appendix \ref{AppendixNoether}, we derive $\Theta^{\mu\nu}$ through Noether's second theorem just by assuming Poincar\`e invariance, in absence of any particular aim. In fact, this suggests that it is the improved tensor which should be considered as the actual canonical current.

\subsection{Interpretation}
Looking at this new energy-momentum tensor in (\ref{BelinTensor}), we identify the first $-$canonical$-$ term with energy and linear-momentum flow across hyperplanes, as usual. The second term, which is of the form\footnote{This form becomes manifest if we consider the density of the associated spin pseudovector $S^\mu=\frac{1}{2}\epsilon^{\mu\nu\rho\sigma}u_\nu S_{\rho\sigma}$, where $u^\mu$ is the four-velocity. We will introduce this vector in the next section.} $\nabla\cross\mathcal{S}$, reflects the breakdown of a spacetime-independent angular-momentum across the same hyperplanes. However, this density does not contribute to the total energy and momentum, which means that the net flow of this kind of angular momentum is exactly zero. Therefore, it must be made of bound currents (i.e. closed circuits) of momentum density, whose existence is irrespective of spacetime position. This is spin, the kind of angular momentum that is \textit{intrinsic} to entities in field theory.\\

Two comments are in place here. Firstly, spin, of course, exists without considering the improved energy-momentum tensor $\Theta^{\mu\nu}$ or even Poincar\`e invariance itself, as seen from considering Noether's theorem solely for Lorentz invariance; it is just the improved picture that provides its nice interpretation as an angular momentum due to (bound) rotational motion. Ohanian \cite{Ohanian:1986} was the first to illustrate this for the electromagnetic and the Dirac field. Secondly, realizing spin as a bound current of angular-momentum density, i.e. $\nabla\cross\mathcal{S}$, is completely analogous to magnetism. There, except the free currents we also have the bound currents $\vec{\mathbf{M}}_b=\nabla\cross\vec{\mathbf{M}}$ (associated with some magnetization density $\vec{\mathbf{M}}$), both of them sourcing the magnetic field. The same applies here, where except the free (canonical) current of four-momentum we include the bound current, which is spin angular momentum.\\

In the presence of a gravitational field, $\Theta^{\mu\nu}$ is identified \cite{Gotay:1992} with the (metric) Hilbert energy-momentum tensor $\mathcal{T_{\tiny H}}^{\mu\nu}=2\frac{\partial\mathcal{L}}{\partial g_{\mu\nu}}+g^{\mu\nu}\mathcal{L}$. Moreover, since, at the quantum level, equations of motion are satisfied in correlators up to contact terms, the energy-momentum tensor redefinition modifies the Ward identities in those contact terms exclusively.\\

Overall, symmetry seems to naturally divide angular momentum into two complementary parts: orbital and spin angular momentum. The first reflects motion inside the target space and is thus of kinematic character, while the second involves some kind of internal (rotational) motion and is thus dynamic w.r.t., e.g., external spacelike fields.\\

\section{Spin-orbit duality}\label{SectionDuality}
In that sense, this kinematic-dynamic complementarity should also have a geometric point of view. A simple take, in that respect, is that kinematics in the theory may as well be defined as the projection along a kinematic four-vector, e.g. the four-momentum. Dynamics should be whatever remains from that projection.

\subsection{$(1+3)$ decomposition}
In four-dimensional spacetime, this projection is the $(1+3)$ decomposition of tensors. With respect to a timelike four-momentum $p^\mu$, where $p^2=-m^2$, we may decompose the flat metric into a parallel and a normal tensor part, i.e. $\eta^{\mu\nu}=p^\mu p^\nu/p^2+h^{\mu\nu}(p)$, that naturally defines the projection tensor

\begin{equation}
h^{\mu\nu}(p)\;\equiv\;\eta^{\mu\nu}-\frac{p^\mu p^\nu}{p^2}
\end{equation}\\
which projects normal to $p^\mu$. Then, a general second-order tensor $X$ in flat space can be $(1+3)$-decomposed as

\begin{equation}
X_{\mu\nu}\;=\;\frac{1}{p^4}(X_{\rho\sigma}p^\rho p^\sigma)p_\mu p_\nu\,+\,\frac{1}{p^2}({h_\mu}^\rho X_{\rho\sigma}p^\sigma)p_\nu\,+\,\frac{1}{p^2}({h_\nu}^\sigma X_{\rho\sigma}p^\rho)p_\mu\,+\,{h_\mu}^\rho{h_\nu}^\sigma X_{\rho\sigma}\;.
\end{equation}\\
When, in particular, $X$ is a bivector, then $X_{\mu\nu}p^\mu p^\nu=0$. In this case, we may define the vector

\begin{equation}
E_\mu\;=\;{h_\mu}^\rho X_{\rho\nu}\,p^\nu\;=\;X_{\mu\nu}\,p^\nu\;,\label{BivectorElectricPart}
\end{equation}\\
and, also, since $({h_\mu}^\rho{h_\nu}^\sigma X_{\rho\sigma})p^\mu=0$, another one as

\begin{equation}
{h_\mu}^\rho{h_\nu}^\sigma X_{\rho\sigma}\;=\;\frac{1}{p^2}\epsilon_{\mu\nu\rho\sigma}p^\rho H^\sigma\hspace{1cm}\Rightarrow\hspace{1cm}H_\mu\;=\;p^\nu\star X_{\mu\nu}\;,\label{BivectorMagneticPart}
\end{equation}\\
where $\star X_{\mu\nu}=\frac{1}{2}\epsilon_{\mu\nu\rho\sigma}X^{\rho\sigma}$ is the Hodge-dual tensor, so that $X$ may be expressed as

\begin{equation}
X_{\mu\nu}\;=\;\frac{1}{p^2}\bigg(E_\mu p_\nu\,-\,E_\nu p_\mu\,+\,\epsilon_{\mu\nu\rho\sigma}p^\rho H^\sigma\bigg)\;.\label{BivectorDecomposition}
\end{equation}\\
We call $E_\mu$ and $H_\mu$ the \textit{electric} and \textit{magnetic part} of $X$, respectively, since, at the rest frame, $X$ takes the familiar form of the electromagnetic tensor, with $E_i$ playing the role of the electric field and $H_i$ that of the magnetic field. Conversely, the traditional electric and magnetic fields may be realized as the distinguished tensor parts of the electromagnetic tensor under the $(1+3)$ decomposition.

Naming

\begin{equation}
\frac{E_\mu p_\nu-E_\nu p_\mu}{p^2}\;\equiv\; E_{\mu\nu}\;,\hspace{2cm}\frac{\epsilon_{\mu\nu\rho\sigma}p^\rho H^\sigma}{p^2}\;\equiv\; H_{\mu\nu}\;,
\end{equation}\\
such that $X_{\mu\nu}=E_{\mu\nu}+H_{\mu\nu}$, we observe that, by construction,
\begin{equation}
p^\mu\star E_{\mu\nu}\;=\;0\hspace{1cm}\mbox{and}\hspace{1cm}H_{\mu\nu}\,p^\nu\;=\;0\;,\label{HodgeDecomposition}
\end{equation}\\
which actually resembles the Helmholtz decomposition of vectors in $\mathbb{R}^3$ into curl-free and divergence-free parts, respectively. In fact, generalizing from vectors to differential forms on (pseudo-)Riemann manifolds, this $(1+3)$ decomposition is a Hodge decomposition. $E_{\mu\nu}$ is the projection of the bivector $X_{\mu\nu}$ along the four-momentum $p^\mu$, while $H_{\mu\nu}$ is the complementary normal projection. So, more or less, $E_{\mu\nu}$ and $H_{\mu\nu}$ may be understood to represent the kinematic and dynamic character of $X_{\mu\nu}$, respectively.\\

Angular momentum is a bivector and, hence, it can be $(1+3)$-decomposed, accordingly. In that sense, since orbital and spin angular momentum reflect too a kind of kinematic-dynamic complementarity, we may wonder whether they fit into the structure of the Hodge decomposition. If they do, then the orbital part should be $L_{\mu\nu}=E_{\mu\nu}$ and the spin part $S_{\mu\nu}=H_{\mu\nu}$, satisfying the Hodge structure (\ref{HodgeDecomposition}). Indeed, this is exactly what happens.

\subsection{Angular-momentum decomposition}
Assuming a certain massive field configuration of four-momentum $p^\mu$, we decompose w.r.t. this four-momentum, the magnetic part of total angular momentum reading

\begin{equation}
H_\mu\;=\;\int_{\mathbb{R}^3}p^\nu\star\mathcal{M}_{\mu\nu}\;=\;p^\nu\star S_{\mu\nu}\;,\label{SmagneticPart}
\end{equation}\\
which involves only spin, since $p^\nu\star\mathcal{L}_{\mu\nu}=0$ identically, using the momentum-space generator $\mathbf{L}_{\mu\nu}=x_\mu p_\nu-x_\nu p_\mu$. Of course, this integrates over three-space to yield the condition

\begin{equation}
p^\nu\star L_{\mu\nu}\;=\;0\;,\label{LmagneticPart}
\end{equation}\\
in terms of total quantities, i.e. that the magnetic part of orbital angular momentum vanishes. In turn, the electric part reads

\begin{equation}
E_\mu\;=\;M_{\mu\nu}p^\nu\;=\;\left(L_{\mu\nu}p^\nu+S_{\mu\nu}p^\nu\right)\;,
\end{equation}\\
where only the orbital part is, in general, non-zero. The spin electric part is obscure, in the sense that we cannot immediately tell whether it vanishes or not. On the other hand, according to the idea that spin and orbital angular momentum are complementary, and since the magnetic part of the latter vanishes identically, we would expect that the spin electric part also vanishes, $S_{\mu\nu}p^\nu=0$. Indeed, since $S_{\mu\nu}S^{\mu\nu}$ for a field (or fields) $q$ is an invariant, then its derivative w.r.t. any affine parameter $\tau$ should vanish,

\begin{equation}
0\;=\;\frac{\dd}{\dd\tau}\left(S_{\mu\nu}S^{\mu\nu}\right)\;=\;2\,S_{\mu\nu}\dot{S}^{\mu\nu}\;=\;-4\,S_{\mu\nu}\int_{\mathbb{R}^3}\frac{\partial\mathcal{L}}{\partial\dot{q}}\left(\frac{\partial\mathcal{L}}{\partial q}\,x^\mu p^\nu q\,+\,\dot{x}^\mu p^\nu q\,+\,x^\mu p^\nu \dot{q}\right)\;,\label{VanishingDevS2}
\end{equation}\\
where we used the angular-momentum conservation law (\ref{TotalAngMomCons2}), $\dot{S}^{\mu\nu}=-\dot{L}^{\mu\nu}$, translation invariance, $\dot{p}^\mu=0$, and the Lagrange equations of motion. Next, we note that the angular-momentum conservation law, (except when $\dot{S}^{\mu\nu}=\dot{L}^{\mu\nu}=0$), states

\begin{equation}
-\dot{S}^{\mu\nu}\;=\;\dot{L}^{\mu\nu}\;\neq\;0\;,
\end{equation}\\
which implies that four-velocity $u^\mu\equiv\dot{x}^\mu$ and four-momentum are not generally parallel. Hence, the second term in (\ref{VanishingDevS2}) does not vanish identically. As a matter of fact, none of the three terms in that equation is a priori zero. Therefore, it must be that

\begin{equation}
S_{\mu\nu}\,p^\nu\;=\;0\;.\label{SelectricPart}
\end{equation}\\
Equivalently, given that this implies $(S_{0\mu})_+=0$ in the rest frame and that the spin tensor has six antisymmetric components in total, we could have just employed Lorentz symmetry to achieve all this. This would look like an arbitrary choice, but this is not so since it has been shown \cite{Pryce:1948,Beigl:1967} that this condition defines a unique center-of-mass worldline. In fact, this constraint goes by the name \textit{`supplementary spin condition'} and has been derived in various ways. One that is similar to the above can be found in \cite{Barandes:2019oas}. The oldest proof dates back to Pryce \cite{Pryce:1948} and Beiglb\"ock \cite{Beigl:1967}, in an effort to describe the relativistic center of mass. Subsequent works reach the same result \cite{Papa:1951}, while a review of the various points of view can be found in \cite{Costa:2014nta}. Note, also, that since most of those proofs do not assume translation invariance, they are more general than ours and, hence, (\ref{SelectricPart}) holds just in the presence of Lorentz symmetry.\\

Therefore, orbital and spin angular momentum decouple geometrically (but not dynamically!), in the sense that they are independent parts of a unified algebraic structure,
\begin{equation}
\left\lbrace\;\;p^\nu{\star L_{\mu\nu}}\;=\;0\;\;,\;\;S_{\mu\nu} p^\nu\;=\;0\;\;\right\rbrace\;,\label{STRUCTURE}
\end{equation}
which, as shown, is a Hodge decomposition. To get a better grip on the decoupling of angular momentum under this decomposition, we notice that, in the rest frame $\Sigma_+$, equations in (\ref{STRUCTURE}) imply $(L_{ij})_+=(S_{0i})_+=0$, so that
\begin{equation}
M_{\mu\nu}\;=\;\left(\begin{array}{cc}
0 & L_{0j}\\
L_{i0} & S_{ij}
\end{array}\right)_+\;.\label{COMtotAngMom}
\end{equation}\\
In this picture, boosts are generated solely by orbital angular momentum and spatial rotations by spin. Or, in the words of relativistic kinematics, in the rest frame, the particle only boosts along spatial dimensions and sees a spacelike field, which is its spin angular momentum.

\subsection{Spin condition}
The use of the constraint $S_{\mu\nu}p^\nu=0$ throughout the diverse literature is, usually, to eliminate unphysical degrees of freedom. Here, it does more than that. That is, as with (\ref{BivectorMagneticPart}), it implies that the spin tensor can be deconstructed as

\begin{equation}
S_{\mu\nu}\;=\;-\frac{1}{p^2}\,\epsilon_{\mu\nu\rho\sigma}W^\rho p^\sigma\;,
\end{equation}\\
where $W^\mu$ makes sense here only as a spacelike vector, i.e. $W^\mu p_\mu=0$. The factor $-1/p^2$ is just a convenience, for this expression to be reversed as

\begin{equation}
W_\mu\;=\;\frac{1}{2}\epsilon_{\mu\nu\rho\sigma}S^{\nu\rho}p^\sigma\;,\hspace{2cm}W_\mu\,p^\mu=0\;.\label{PauliLubanskiVector}
\end{equation}\\
$W^\mu$ is, in fact, proportional to a spin pseudovector, $S_\mu=\frac{1}{2}\epsilon_{\mu\nu\rho\sigma}S^{\nu\rho}u^\sigma$, the two vectors related, in the rest frame, as $W_\mu=(0,mS_i)_+$. It is, also, easy to check that $\frac{1}{2}S^{\mu\nu}S_{\mu\nu}=S^\mu S_\mu\equiv S^2$ and, more importantly,

\begin{equation}
W^2\;=\;m^2S^2\;.
\end{equation}\\
In fact, we have just reached the contact point between the classical and quantum theory; the rest-frame vector $S^i$ and the invariant $S^2$, lifted as quantum operators acting on states in a Hilbert space, define the eigenvalue problem which makes sense out of the usual particle spin.\\

Moreover, we notice that $W^\mu$ is exactly the magnetic part of the angular momentum tensor, (\ref{SmagneticPart}); actually, spin being packaged into the magnetic part $H^\mu=W^\mu$ of total angular momentum is the the whole point of the Hodge decomposition. Of course, $W^\mu$ is not just any vector; it is the famous Pauli-Lubanski pseudovector. In the level of the algebra, in the rest frame, we have $\mathbf{W}_\mu=(0,m\mathbf{S}_i)_+$, implying $[\mathbf{W}_i,\mathbf{W}_j]=im\,\epsilon_{ijk}\mathbf{W}^k$, which is an enveloping algebra of $[\mathbf{S}_i,\mathbf{S}_j]=i\,\epsilon_{ijk}\mathbf{S}^k$ that defines the SO$(3)$ little group of SO$(1,3)$. $\mathbf{W}^\mu\mathbf{W}_\mu$, along with $\mathbf{P}^\mu\mathbf{P}_\mu$, are the unique Casimir elements of the Poincar\`e algebra. This is why all irreducible unitary representations of the Poincar\`e group are labeled exclusively by their mass and spin. In a quantum theory of massive states $\ket{m,s}$, of mass $m$ and spin $s$, since $\hat{W}^2=m^2\hat{S}^2$, we have the eigenvalues $\hat{W}^2\ket{m,s}=m^2\hbar^2s(s+1)\ket{m,s}$. Eventually, the natural appearance of the Pauli-Lubanski pseudovector in the Hodge decomposition of angular momentum as its magnetic part may serve as another definition of this vector.

\subsection{An electric-magnetic duality}
The statement that spin and orbit decouple geometrically under the Hodge decomposition is just an expression for the structure (\ref{STRUCTURE}), $\lbrace p^\nu{\star L_{\mu\nu}}=0\,, S_{\mu\nu} p^\nu=0\rbrace$. For the mind that always seeks symmetry, a natural observation here is that this structure is invariant under the exchange between the magnetic and electric part of angular momentum, $H_\mu=p^\nu{\star S_{\mu\nu}}$ and $E_\mu=L_{\mu\nu}p^\nu$. That is,

\begin{equation}
L^{\mu\nu}\,\mapsto\,\star{S}^{\mu\nu}\;,\hspace{1cm}S^{\mu\nu}\,\mapsto\,\star{L}^{\mu\nu}\;,\label{SLinterchange}
\end{equation}\\
is an automorphism of this structure. This symmetry raises the question of whether this automorphism yields a meaningful duality of the theory as a whole. Another strong motive to look into this is that Hodge decomposition of angular momentum poses a striking resemblance with electromagnetism, which enjoys the electric-magnetic duality; this is really the exchange between the associated electric and magnetic parts of the electromagnetic field-strength bivector. The resemblance is especially manifest in the rest-frame representation of (\ref{COMtotAngMom}). From this perspective, it is more than natural to ask whether such a duality between orbital and spin angular momentum represents a meaningful concept.\\

Electric-magnetic duality is the statement that Maxwell's equations are invariant under the exchange of fields $(\vec{E},\vec{B})\mapsto(\vec{B},-\vec{E})$. Equivalently, in U$(1)$ pure gauge theory, where Lorentz invariance is manifest, the equations of motion (and Bianchi identity),

\begin{equation}
\partial_\mu\,F^{\mu\nu}\;=\;0\;,\hspace{1cm}\partial_\mu{\star{F}}^{\mu\nu}\;=\;0\;,
\end{equation}\\
where ${\star F}^{\mu\nu}=\frac{1}{2}\epsilon^{\mu\nu\rho\sigma}F_{\rho\sigma}$ is the Hodge-dual strength, are invariant under the map
\begin{equation}
F^{\mu\nu}\,\mapsto\,\star{F}^{\mu\nu}\;,\hspace{1cm}{\star F}^{\mu\nu}\,\mapsto\,-F^{\mu\nu}\;.
\end{equation}
The negative sign in the second case is because $\star^2=-1$ in a four-dimensional Lorentz manifold. An associated map holds between electric and magnetic source currents, if those are present in the theory, and overall this duality is an honest symmetry of electromagnetism. In that respect, there is a sensible analogy with angular momentum,

\begin{equation}
M^{\mu\nu}\,\mapsto\,\star{M}^{\mu\nu}\;,\hspace{1cm}{\star M}^{\mu\nu}\,\mapsto\,-M^{\mu\nu}\;,\label{Minterchange}
\end{equation}
since, given Hodge decomposition, this actually implies the map $\lbrace L^{\mu\nu}\mapsto\star{S}^{\mu\nu}\,, S^{\mu\nu}\mapsto\star{L}^{\mu\nu}\rbrace$\footnote{In fact, the duality map is true up to a sign, i.e. $M^{\mu\nu}\,\mapsto\,\pm(\star{M}^{\mu\nu})$ etc. Equivalently, $L^{\mu\nu}\,\mapsto\,\pm(\star{S}^{\mu\nu})\,$ and $S^{\mu\nu}\,\mapsto\,\pm(\star{L}^{\mu\nu})$. We adopt the positive sign as our convention, without any loss of generality. This choice produces orientation-related minus signs throughout this article, which however are physically unimportant.}, which is the suggested structural symmetry (\ref{SLinterchange}). We shall call this the spin-orbit duality. Again, this map leaves the geometric structure (\ref{STRUCTURE}) intact,
\begin{equation}
\begin{array}{ccc}
p^\nu{\star L}_{\mu\nu}=0\hspace{0.5cm} & \mapsto\hspace{0.5cm} & S_{\mu\nu}p^\nu=0\\[10pt]
S_{\mu\nu}p^\nu=0\hspace{0.5cm} & \mapsto\hspace{0.5cm} & p^\nu{\star L}_{\mu\nu}=0\;,
\end{array}
\end{equation}\\
which was the original motivation to investigate this duality, in the first place. Under this map, conservation of total angular momentum,

\begin{equation}
\dot{M}^{\mu\nu}\;=\;\dot{L}^{\mu\nu}+\dot{S}^{\mu\nu}\;=\;0\,\hspace{0.5cm}\mapsto\hspace{0.5cm}\,\star{\dot{M}}^{\mu\nu}\;=\;0\;\Leftrightarrow\;\dot{M}^{\mu\nu}\;=\;0\;,
\end{equation}\\
remains invariant, which indicates that this could indeed be a symmetry of the associated conservation law. In turn, in order to test the conservation law of the associated currents, in accordance with the map (\ref{SLinterchange}), we consider the (more fundamental) map

\begin{equation}
\mathbf{L}^{\mu\nu}\,\mapsto\,\star{\mathbf{S}}^{\mu\nu}\;,\hspace{1cm}\mathbf{S}^{\mu\nu}\,\mapsto\,\star{\mathbf{L}}^{\mu\nu}\;,\label{DualLorentzGens}
\end{equation}\\
which exchanges the representations of Lorentz generators or, equivalently, maps $\mathbf{M}\mapsto{\star\mathbf{M}}$. Here, of course, the action of those generators on fields is understood. It is easy to see that this results in a map for densities,

\begin{equation}
\mathcal{L}^{\mu\nu}\,\mapsto\,\star{\mathcal{S}}^{\mu\nu}\;,\hspace{1cm}\mathcal{S}^{\mu\nu}\,\mapsto\,\star{\mathcal{L}}^{\mu\nu}\;,\label{DensityDuality}
\end{equation}\\
and current densities,

\begin{equation}
\mathcal{L}^{\rho\mu\nu}\,\mapsto\,\frac{1}{2}\epsilon^{\mu\nu\alpha\beta}{\mathcal{S}^\rho}_{\alpha\beta}\;,\hspace{1cm}\mathcal{S}^{\rho\mu\nu}\,\mapsto\,\frac{1}{2}\epsilon^{\mu\nu\alpha\beta}{\mathcal{L}^\rho}_{\alpha\beta}\;,\label{CurrentDuality}
\end{equation}\\
under which the general conservation law,

\begin{equation}
\partial_\rho\mathcal{M}^{\rho\mu\nu}\;=\;\partial_\rho\left(\mathcal{L}^{\rho\mu\nu}+\mathcal{S}^{\rho\mu\nu}\right)\;=\;0\,\hspace{0.5cm}\mapsto\hspace{0.5cm}\,\epsilon^{\mu\nu\alpha\beta}\partial_\rho{\mathcal{M}^\rho}_{\alpha\beta}\;=\;0\;\Leftrightarrow\;\partial_\rho\mathcal{M}^{\rho\mu\nu}\;=\;0\;,\label{AngMomCurrConsSOduality}
\end{equation}\\
remains invariant. Therefore, although interchanging internal with external degrees of freedom at best seems unconventional, spin-orbit duality respects Lorentz symmetry. In fact, this is just the inverse Noether's theorem: conserved quantities coming from conservation laws are generators of infinitesimal symmetry transformations\footnote{To be precise, there is a couple of obvious assumptions to this inversion. The first is that our system should be Hamiltonian, which it is. The second assumption is that the conserved quantities are non-degenerate, which, at least in this case, they are. This is, of course, the case, since the dual conserved quantities are, as a final set, the same as the ones in the initial theory: the six components of total angular momentum.}. Hence, since angular momentum is conserved in the dual theory too, it still corresponds to a continuous symmetry.

\subsubsection*{Invariance of the Lorentz Algebra}
In the level of the algebra, as expected from preservation of Lorentz symmetry, the structure of the $\mathfrak{so}(1,3)$ algebra is invariant under the spin-orbit duality. That is, schematically,

\begin{equation}
[\mathbf{M},\mathbf{M}]=\eta\mathbf{M}+\eta\mathbf{M}-\eta\mathbf{M}-\eta\mathbf{M}\hspace{1cm}\mapsto\hspace{1cm}[\widetilde{\mathbf{M}},\widetilde{\mathbf{M}}]=\eta\widetilde{\mathbf{M}}+\eta\widetilde{\mathbf{M}}-\eta\widetilde{\mathbf{M}}-\eta\widetilde{\mathbf{M}}\;\;,\label{DualLorentzAlgebra}
\end{equation}
where $\widetilde{\mathbf{M}}=\star\mathbf{M}$. Nonetheless, it is instructive to see the dual algebra through the eyes of the initial theory. That is, we replace for the original generators in the dual algebra, $\widetilde{\mathbf{M}}=\star\mathbf{M}$,

\begin{equation}
[\mathbf{M}_{\mu\nu},\mathbf{M}_{\rho\sigma}]=\eta^{\alpha\beta}\epsilon_{\beta\nu\rho\sigma}\mathbf{M}_{\alpha\mu}-\eta^{\alpha\beta}\epsilon_{\beta\mu\rho\sigma}\mathbf{M}_{\alpha\nu}\;.\label{DualAlgebraOriginalGens}
\end{equation}\\
This is the Lorentz algebra in the dual theory, but in terms of the generators of the initial theory. Therefore, in terms of the original generators, the duality map implies, for example,
\begin{equation}
\begin{split}
[\mathbf{M}_{01},\mathbf{M}_{02}]\;=\;-\mathbf{M}_{12}\hspace{1cm}&\mapsto\hspace{1cm}[\mathbf{M}_{01},\mathbf{M}_{02}]\;=\;\mathbf{M}_{30}\;,\\[10pt]
[\mathbf{M}_{01},\mathbf{M}_{23}]\;=\;0\hspace{1cm}&\mapsto\hspace{1cm}[\mathbf{M}_{01},\mathbf{M}_{23}]\;=\;0\;,
\end{split}\label{DualAlgebraOriginalGensExample}
\end{equation}
where same commutators project onto Hodge-dual generators across the dual pictures, up to an orientation-related change of sign. Again, this twisted dual picture does certainly \textit{not} mean that Lorentz symmetry is lost; spin-orbit duality preserves the conservation of the angular-momentum current, which yields that Lorentz symmetry must still be present. What happens is that Hodge duality, $\mathbf{M}\mapsto\star\mathbf{M}$, shuffles the generators (not changing them per se) and so their Lorentz algebra shifts its basis. Indeed, the Hodge star operator is a linear map that just takes an orthonormal basis into an orthonormal basis. Hence, Hodge duality is an automorphism of their algebra. This is also reflected in the two Casimir elements of the Lorentz algebra which are left invariant, again up to a change of sign,
\begin{equation}
\begin{split}
\mathbf{M}^2\hspace{1cm}&\mapsto\hspace{1cm}-\mathbf{M}^2\\[10pt]
{\star\mathbf{M}}\mathbf{M}\hspace{1cm}&\mapsto\hspace{1cm}-({\star\mathbf{M}}\mathbf{M})\;.
\end{split}
\end{equation}
An equivalent, more geometric way to see that this an automorphism is to consider the constrained Lorentz group SO$^+(1,3)$ as a manifold. Since boosts are the homogeneous space SO$^+(1,3)/$SO$(3)\cong\mathsf{H}^3\cong\mathsf{R}^3$\; and rotations are \;SO$(3)\cong\mathsf{RP}^3$, then \;SO$^+(1,3)\cong\mathsf{RP}^3\cross\mathsf{R}^3$. In terms of the SO$^+(1,3)$ group manifold, spin-orbit duality $\mathbf{M}\mapsto\star\mathbf{M}$, which is an exchange between boost and spatial-rotation generators, is realized as an (orthonormal) swapping between timelike planes $\mathbf{M}_{0i}$ and spacelike planes $\mathbf{M}_{jk}$. Therefore, the duality is an isomorphism between spaces $\mathsf{RP}^3$ and $\mathsf{R}^3$, implying an overall isomorphism
\begin{equation}
\mathsf{RP}^3\cross\mathsf{R}^3\hspace{1cm}\mapsto\hspace{1cm}\mathsf{R}^3\cross\mathsf{RP}^3\;,\label{LorentzManifoldDuality}
\end{equation}
which leaves the product space $\mathsf{RP}^3\cross\mathsf{R}^3$ topologically invariant.

In the case of a Lorentz-invariant theory, conservation of angular momentum under the spin-orbit duality, i.e. the invariance (\ref{AngMomCurrConsSOduality}), is the only conservation law whose invariance we have to investigate. In a Poincar\`e-invariant theory, however, we also ought to study whether the $\mathfrak{iso}(1,3)$ algebra and conservation of energy-momentum remain invariant. In our case, we do have to do this, because the duality acts on orbital and spin angular momenta, which depend explicitly on the energy-momentum density, i.e. $\partial_\rho\,\mathcal{L}^{\rho\mu\nu}=-\partial_\rho\,\mathcal{S}^{\rho\mu\nu}=\mathcal{T}^{\mu\nu}-\mathcal{T}^{\nu\mu}$, and, thus, conservation of the energy-momentum current is at stake. As it turns out, energy and momentum are still conserved in the dual picture but, this time, the proof is a bit more elaborate and, hence, we present all its details in Appendix \ref{AppendixEMTsusy}. The bottom line is that translations are still a symmetry in the dual theory and this duality is enjoyed by the more-constrained class of Poincar\`e-invariant massive field theories.

\subsection{Dual momentum and position}\label{SubsectionDualMomPos}
So we have seen the way angular momenta transform under the proposed duality, which was the very inspiration for this duality to begin with. On the other hand, orbital angular momentum itself is really the exterior algebra of position and momentum and, thus, a more fundamental map must be implied for those quantities.\\

Taking up first the simplest of the two, four-momentum should stay a four-momentum. This is not a physical assumption, nor an axiom of any sort, but a consequence of the nature of the spin-orbit duality. Let us explain. If we call $\mathcal{H}:=(\mathbf{S},\mathbf{L},\star)=(p^\nu{\star L_{\mu\nu}}=0\,, S_{\mu\nu} p^\nu=0)$ the (Hodge) structure on the cotangent bundle $T^*M$, then the duality may be defined as an endomorphism $F$ on $\mathcal{H}$,
\begin{equation}
F:\;\;\mathcal{H}\hspace{0.3cm}\mapsto\hspace{0.3cm}\mathcal{H}\;,
\end{equation}
which is also an isomorphism (since it is bijective) and, hence, an automorphism. If four-momentum transforms under $F$ into a general four-vector $\tilde{p}^\mu$, then the map of $\mathcal{H}$ is

\begin{equation}
0\;=\;S_{\mu\nu}p^\nu\hspace{0.5cm}\mapsto\hspace{0.5cm}\tilde{p}^\nu{\star L_{\mu\nu}}\;\stackrel{!}{=}\;0\;.
\end{equation}\\
This remains an automorphism, if and only if $\tilde{p}^\mu\propto p^\mu$. The simplest example is dilations $\tilde{p}^\mu=\lambda p^\mu$, $\lambda\in\mathbb{R}$. In the general case, though, it may be $\tilde{p}^\mu=f p^\mu$, with $f$ a scalar function of a general argument, which, by the way, also respects the form of the projection tensor $h^{\mu\nu}(p)=g^{\mu\nu}-p^\mu p^\nu/p^2$. This last remark is equally important, since $\mathcal{H}$ is actually a Hodge decomposition, which, in this context, is a $(1+3)$ decomposition, which, in turn, relies entirely on $h^{\mu\nu}(p)$. In any case, as we are about to see, $p^\mu$ and $x^\mu$ are entangled under the map $F: L_{\mu\nu}\mapsto{\star S_{\mu\nu}}$ and, thus, a transformation $\tilde{p}^\mu=f p^\mu$ would just parametrize $x^\mu\mapsto\frac{1}{f}\tilde{x}^\mu$, whatever $\tilde{x}^\mu$ is. Hence, without loss of generality, we may as well choose $f=1$ and set
\begin{equation}
F:\;\;{p}^\mu\hspace{0.3cm}\mapsto\hspace{0.3cm}p^\mu\;,
\end{equation}
to make everything prettier. Hence, four-momentum stays the same under the duality.

Of course, as with angular momentum in (\ref{DualLorentzGens}), this is taken to imply a more fundamental map, that of the translation generators, $\mathbf{P}_\mu\mapsto\mathbf{P}_\mu$. Therefore, it is now straightforward to check that the Poincar\`e algebra remains intact under the spin-orbit duality, something that should have already been anticipated from the preservation of the energy-momentum current in the dual theory. The proof is analogous to that of the invariance of the Lorentz algebra and is presented in Appendix \ref{AppendixPoincareAlgebra}.\\

To find how four-position transforms, we consider the original map between $L^{\mu\nu}\mapsto\star S^{\mu\nu}$. In fact, we actually pick the corresponding map between densities, $\mathcal{L}^{\mu\nu}\mapsto\star\mathcal{S}^{\mu\nu}$, or the one between generators, $x^\mu p^\nu-x^\nu p^\mu=\mathbf{L}^{\mu\nu}\mapsto\,\star \mathbf{S}^{\mu\nu}$ (where their action on fields is understood), which, given that four-momentum does not change, yield

\begin{equation}
h^\mu\hspace{0.5cm}\mapsto\hspace{0.5cm}\tilde{h}^\mu\;\equiv\;\frac{W^\mu}{p^2}\;.\label{DualPositionSpatial}
\end{equation}\\
Here, $h^\mu\equiv{h^\mu}_\nu(p)\,x^\nu$ is the spacelike part of the four-position, the latter being $(1+3)$-decomposed w.r.t. four-momentum as

\begin{equation}
x^\mu\;=\;\frac{(x\cdot p)\,p^\mu}{p^2}\,+\,{h^\mu}_\nu(p)x^\nu\;\equiv\;y^\mu\,+\,h^\mu\;,\label{Xdecomposition}
\end{equation}\\
where we also defined the timelike part $y^\mu\equiv(x\cdot p)p^\mu/p^2$. Accordingly, $\;S^{\mu\nu}\mapsto\star L^{\mu\nu}\;$ leads to $\;W^\mu\mapsto\tilde{W}^\mu\equiv -p^2\,{h^\mu}$. Hence, the dual spacelike position $\tilde{h}^\mu$ is defined by the Pauli-Lubanski vector $W^\mu$ (i.e. the spin and mass) of the initial theory and, conversely, the dual Pauli-Lubanski vector $\tilde{W}^\mu$ is defined by the position $h^\mu$ of the initial theory. Had we let for a transformation $p^\mu\mapsto\tilde{p}^\mu=f p^\mu$, then the right-hand side of the map (\ref{DualPositionSpatial}) would simply acquire a factor of $1/f$, which makes obvious our freedom to conveniently set $f=1$. In any case, since $W_\mu p^\mu=0$, this is a map between spacelike objects, which in the rest frame reads

\begin{equation}
x^i\;(\;=h^i)\hspace{0.5cm}\mapsto\hspace{0.5cm}\tilde{x}^i\;(\;=\tilde{h}^i)\;\equiv\;\frac{W^i}{p^2}\;,\label{DualPositionSpatialCOM}
\end{equation}\\
where we did not assign any subscripts since we always refer to the rest frame when we talk about spatial four-vector components.\\

\begin{figure}[t!]
    \centering
    {{\includegraphics[width=10cm]{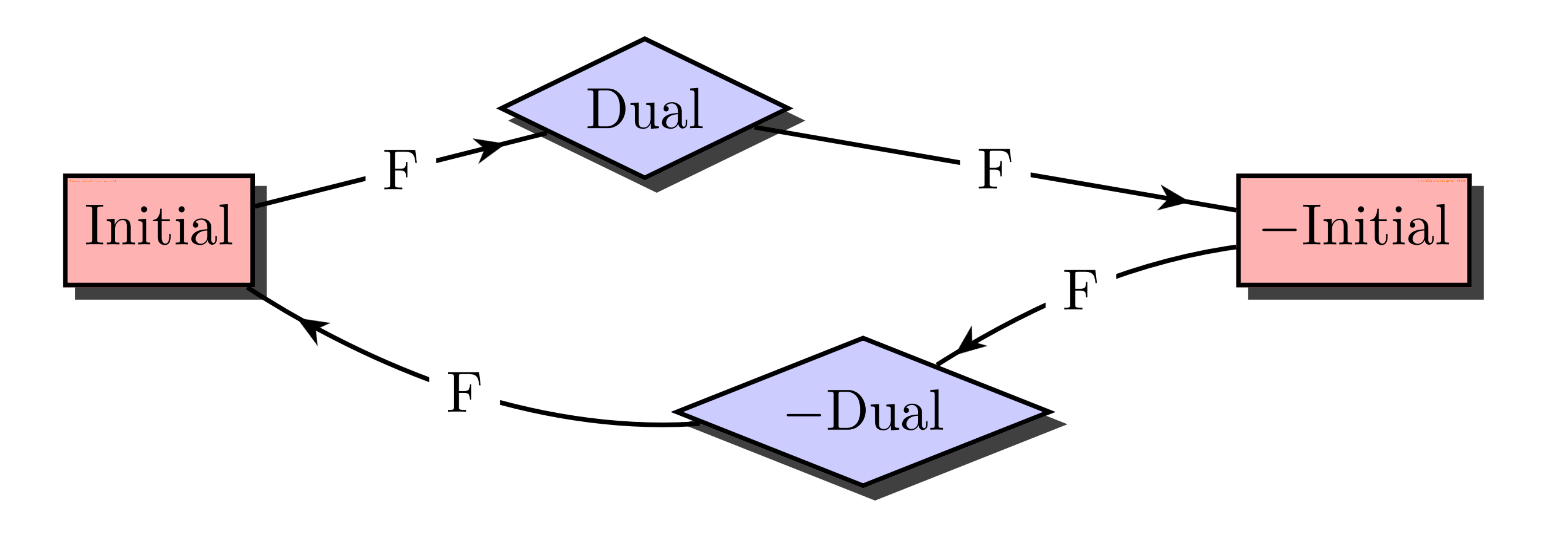} }}%
\caption{Given the spin-orbit duality map $F$, angular momentum needs a quadruple action of $F$ in order to return to its initial form, i.e. $L_{\mu\nu}\mapsto\star S_{\mu\nu}\mapsto-L_{\mu\nu}\mapsto-(\star S_{\mu\nu})\mapsto L_{\mu\nu}$. The same is obvious, through the Lorentz algebra duality map (\ref{DualLorentzAlgebra}). Here, we symbolically add a minus sign in front of the theory, depending on which step of the map sequence it is.}
\label{MAP}
\end{figure}

Exactly because orbital angular momentum is defined on the exterior algebra of position and momentum, the map of the spacelike four-position $h^\mu$ is all the information we can extract from the spin-orbit duality. That is, because it is really $\mathbf{L}_{\mu\nu}=h_\mu p_\nu-h_\nu p_\mu$, then the duality $L^{\mu\nu}\mapsto\star S^{\mu\nu}$ deals only with $h_\mu$ and poses no restriction on the timelike part, $y^\mu$, whatsoever. Therefore, since $p^\mu\mapsto p^\mu$, we may as well take the four-position part that is parallel to the four-momentum to stay the same, without loss of any generality. This is the statement that
\begin{equation}
y^\mu\hspace{0.5cm}\mapsto\hspace{0.5cm}y^\mu\;.
\end{equation}
Thus, for example, in the rest frame, this implies $(x^0)_+\equiv x^0\mapsto x^0$ and the duality is

\begin{equation}
x^\mu\;=\;y^\mu+h^\mu\;=\;\left(\begin{array}{c} x^0\\ 0\end{array}\right)_++\left(\begin{array}{c} 0\\ x^i\end{array}\right)_+\hspace{0.5cm}\mapsto\hspace{0.5cm}\left(\begin{array}{c} x^0\\ 0\end{array}\right)_++\left(\begin{array}{c} 0\\ \frac{W^i}{p^2}\end{array}\right)_+\;=\;\tilde{x}^\mu\;.\label{DualPositionCOM}
\end{equation}\\
In a general frame, we acquire

\begin{equation}
x^\mu\;=\;y^\mu\,+\,h^\mu\hspace{0.5cm}\mapsto\hspace{0.5cm}\tilde{x}^\mu\;=\;y^\mu\,+\,\frac{W^\mu}{p^2}\;,\label{DualPosition}
\end{equation}\\
where the dual position is broken down into timelike and spacelike parts, in a natural manner. It is straightforward to check that, given this map for the four-position, a full circle of the spin-orbit duality back to the initial theory, i.e. $L_{\mu\nu}\mapsto\star S_{\mu\nu}\mapsto-L_{\mu\nu}\mapsto-(\star S_{\mu\nu})\mapsto L_{\mu\nu}$, is satisfied in every step of the way, as it should. In fact, this circle is already obvious through the map of the Lorentz algebra (\ref{DualLorentzAlgebra}), a situation presented in Figure \ref{MAP}.\\

The first interesting fact coming out of the four-position map (\ref{DualPosition}) is that, in a general frame $\Sigma$, the dual spacelike position defines a surface

\begin{equation}
\rho^2\;=\;\tilde{h}^\mu\tilde{h}_\mu\;=\;-(\tilde{h}^0_{\mbox{\tiny$\Sigma$}})^2\,+\,(\tilde{h}^i_{\mbox{\tiny$\Sigma$}})^2\;\label{DualDeSitter},\hspace{2cm}\rho^2\;\equiv\;\frac{W^2}{p^4}\;,
\end{equation}\\
which is a three-dimensional one-sheet hyperboloid in spacetime, a de Sitter world-tube of radius $\rho$. Hence, in the dual theory, the hyperspace normal to four-momentum is a de Sitter space, whose curvature is dictated by the Casimirs of the Poincar\`e algebra. In fact, this is the literal manifestation of the spin-orbit duality where spin and position exchange roles: what was before realized as a (particular) spin vector, in the dual picture is a (particular) spacelike position $\tilde{h}^\mu$ living on a de Sitter surface.

In the rest frame, on the other hand, in view of (\ref{DualPositionCOM}) or even (\ref{DualPositionSpatialCOM}), the dual theory sits exclusively on a two-sphere of radius

\begin{equation}
\tilde{x}^i\tilde{x}_i\;=\;\tilde{h}^i\tilde{h}_i\;=\;\rho^2\;.\label{DualSphere}
\end{equation}\\
As time evolves, this defines a cylindrical world-tube in four-dimensional spacetime. Again, this is a particular surface, since it is rendered of the two Poincar\`e Casimirs, and is, thus, fixed. This radius, actually, turns out to be a very special quantity, both in the context of the spin-orbit duality and in relativistic mechanics. At this stage, though, we just want to emphasize that the spacelike position-space of the dual theory is a world-tube of radius $\rho$\footnote{Note that $h^\mu h_\mu=c^2$, for a real $c$, in the initial theory too. That is, the initial spacelike position sits in a three-dimensional de Sitter submanifold too. However, the situation is different now. First, $c$ may anything in the initial theory, i.e. the center of mass may as well be anywhere in spacetime. Secondly, $c$ is a Lorentz invariant quantity but, in general, not a constant; it may be a function of an affine parameter. Hence, all in all, in the initial theory spacetime foliates into de Sitter subspaces of various radii and $h^\mu$ effectively parametrizes all four-dimensional space. Actually, we will soon discover that this is true up to a minimum radius. Anyway, this contrasts the dual theory where $\tilde{h}^\mu$ lives on a de Sitter world-tube of fixed radius, which yields a duality that effectively maps a four-dimensional region onto a three-dimensional one. That is, much like a hologram.}. \\

At this point, we should make an important remark. Throughout this section, four-position $x^\mu$ could refer to nothing else, other than the center of mass of the field configuration. In fact, as already pointed out,
\begin{equation}
\mathbf{L}_{\mu\nu}\;=\;x_\mu p_\nu\,-\,x_\nu p_\mu\;=\;h_\mu p_\nu-h_\nu p_\mu\;,
\end{equation}\\
so that, actually, it is the spacelike $h^\mu$ that represents the center-of-mass position. This is an important realization, because the very notion of the center-of-mass position is obscure in a relativistic theory \cite{Pauri:1975mr,Lorce:2018zpf}, while it is has long been shown \cite{Pryce:1948,Fokker,Moller} that it is comprised not by one, but by three objects: the canonical but non-covariant Newton-Wigner center of mass, the non-canonical and non-covariant M\o{}ller center of energy and the covariant but non-canonical Fokker-Pryce center of inertia. Those are usually dubbed as the \textit{relativistic collective coordinates}. Here, $h^\mu$ is identified with the Fokker-Pryce four-vector, which makes sense since that is the only collective variable that is covariant. Indeed, the Poisson structure of (\ref{Xdecomposition}) yields a non-canonical $h^\mu$ as expected, a fact that is structural for the relativistic angular momentum. What is the relation of the center of inertia, $h^\mu$, with the (notion of) center of mass, is a question that is naturally answered in the next section.\\

Coming back to the dual four-position $\tilde{x}^\mu$ in (\ref{DualPosition}), we notice that, although dimensionally correct, its peculiar form does not provide any deep insight. At least not classically. Quantum-mechanically, as we will soon find out, it has profound consequences. But in order to see those, there is still one special limit of this duality left to consider.

\subsection{Self-duality}\label{SubsectionSelfDuality}
\begin{figure}[t!]
    \centering
    {{\includegraphics[width=15cm]{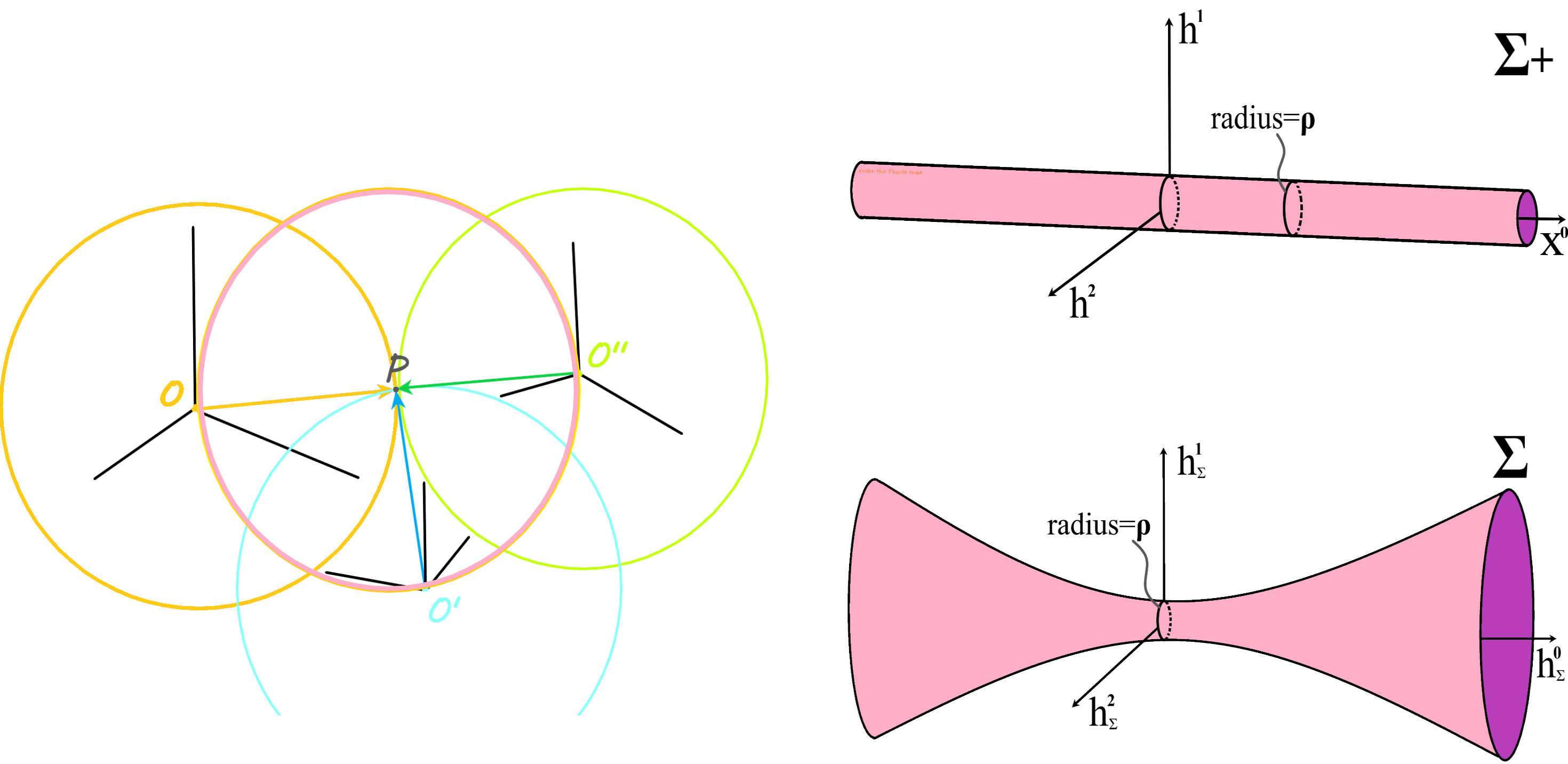} }}%
\caption{On the left: in the rest frame, the center-of-inertia spatial vector $x^i=h^i$ is ambiguous under redefinition of the coordinate origin, here e.g. $(PO)$, $(PO')$ and $(PO'')$. The statement $x^ix_i=h^ih_i=\rho^2$ means that all those vectors have the same length $\rho$. Putting all the (infinite) choices of origins together ($O,O',O'',\ldots$), we obtain a two-sphere (pink sphere) of radius $\rho=S/m$ around the center of inertia $P$. We call this the self-dual two-sphere, on which angular momentum is self-dual and the spin-orbit duality is the trivial map. On the right: the upper graph shows the self-dual flat world-tube, carved by the self-dual two-sphere as time evolves in the rest frame $\Sigma_+$. The lower graph shows the self-dual de Sitter world-tube, in a general frame $\Sigma$. Both are of radius $\rho$. (For de Sitter, $h_{\mbox{\tiny$\Sigma$}}^3=0$ is taken so that the throat at $h_{\mbox{\tiny$\Sigma$}}^0=0$ has radius $\rho$.) Whatever the frame is, the duality maps any $h^\mu$ in the bulk spacetime (white), except the world-tubes' surface (pink) where it becomes trivial.}
\label{moller}
\end{figure}

Since we are dealing with a duality transformation, we may wonder what happens when nothing happens. That is, what is the physical picture when the action of the spin-orbit duality becomes trivial? This is the special occasion when

\begin{equation}
M_{\mu\nu}\;=\;{\star M_{\mu\nu}}\;,
\end{equation}\\
which, of course, since Hodge decomposition decouples spin and orbit, is the statement that

\begin{equation}
L_{\mu\nu}\;=\;\star S_{\mu\nu}\;.
\end{equation}\\
Obviously, this is solved to give the expressions (\ref{DualPositionSpatial})-(\ref{DualPosition}) but this time as equations and not as maps, implying

\begin{equation}
h^\mu\;=\;\tilde{h}^\mu\;=\;\frac{W^\mu}{p^2}\;.
\end{equation}\\
This is understood as follows. While the duality transforms the spacelike four-position $h^\mu$ (which in principle spans all spacetime) into its dual counterpart $\tilde{h}^\mu=W^\mu/p^2$, when we reach the exact value $h^\mu=\tilde{h}^\mu=W^\mu/p^2$ the duality becomes the trivial map. Therefore, actually, for a general frame $\Sigma$, any $h^\mu$ gets mapped except those that live on a \textit{de Sitter world-tube},

\begin{equation}
\rho^2\;=\;h^\mu h_\mu\;=\;-(h^0_{\mbox{\tiny$\Sigma$}})^2\,+\,(h^i_{\mbox{\tiny$\Sigma$}})^2\;,\hspace{2cm}\rho^2\;=\;\frac{W^2}{p^4}\;,
\end{equation}\\
on which angular momentum is self-dual and the spin-orbit duality is trivial. At the same time, this tube represents a natural boundary around the (center of inertia of the) massive configuration. All this is remarkable because, as shown in (\ref{DualDeSitter}), the dual configuration lives on such a surface, which implies that the duality does not only reflect a qualitative exchange between spin and spatial degrees of freedom, but also transforms between \textit{complementary regions} of spacetime. In the rest frame, the same radius,

\begin{equation}
\rho\;=\;\sqrt{x^ix_i}\left(\;=\;\sqrt{h^ih_i}\;=\;\sqrt{\tilde{h}^i\tilde{h}_i}\right)\;=\;\sqrt{\tilde{x}^i\tilde{x}_i}\;=\;\sqrt{\frac{W^2}{p^4}}\;=\;\frac{S}{m}\;,
\end{equation}\\
defines a two-sphere around the center of inertia, while $S=\sqrt{S_iS^i}$ is understood. Eventually, as time evolves, this defines a \textit{flat world-tube} in four-dimensional spacetime; it is this locus $-$the surface of the world-tube$-$ where self-duality of angular momentum occurs. Those induced world-tubes, along with the basic fact that $|\vec{h}|=\rho$ defines a boundary around the massive configuration, are explained in Figure \ref{moller}.\\

But this is a striking result: $\rho$ is the so-called M\o{}ller radius \cite{Moller}. Given that in a relativistic system with nonzero spin there is no canonical four-position that describes the center of mass \cite{Pryce:1948} $-$but only pseudo-worldlines which depend on the chosen inertial frame$-$ M\o{}ller showed that all those pseudo-worldlines fill in a world-tube of an invariant transverse radius $\rho$, exactly like the one we found. The geometric center of the tube is the worldline of the Fokker-Pryce center of inertia, which, in our context, is the spacelike four-position $h^\mu$. This world-tube defines a region of non-covariance of the center-of-mass worldline and imposes a limit on the localization of the canonical center-of-mass. In other words, the sought-for center of mass may be regarded as a fuzzy notion, an abstraction enveloped inside that world-tube \cite{Lorce:2018zpf,Lusanna:1994un}. It has, also, been shown that the same $\rho$ is the minimum radius of a volume that a spinning relativistic body has to have, in order not to rotate with a superluminal velocity. In fact, if the material volume of a rotating system has a radius smaller that than $\rho$, then there are frames where energy is not positive-definite, which renders this radius a reduction of the energy conditions of General Relativity down in flat spacetime \cite{Lusanna:1995bu}.

Quantum-mechanically, M\o{}ller radius is an operator that acts on a massive state $\ket{m,s}$, reflecting the eigenvalue
\begin{equation}
\hat{\rho}\;=\;\frac{\sqrt{s(s+1)}\hbar}{m}\;,
\end{equation}\\
which is of the order of the Compton wavelength, $\lambda_C=\hbar/m$, of the associated field configuration. In other words, the configuration travels through spacetime spanning a world-tube of radius $\rho\approx\lambda_C$, which envelopes its (center-of-mass) position. Therefore, this world-tube not only encloses all frame-dependent (pseudo-)worldlines of the center-of-mass, but is also a region where relativistic (classical) physics seizes to be valid, since localization in such small regions implies pair production; the notion of a single, localized particle breaks down completely below the Compton wavelength. All in all, the existence of the tube reflects, both from a classical and quantum point of view, a limitation in localizing the center-of-mass position. In that respect, as we could have already guessed, the M\o{}ller radius has been suggested as a candidate for an effective ultraviolet (UV) cutoff \cite{Lusanna:1994un}. As are about to see in the next section, such a regularization comes up naturally in the dual quantum theory.\\

Returning to our original cause, the surface of the \textit{self-dual world-tube} of a massive field configuration is, as said, the locus where the duality itself becomes trivial. On the other hand, the dual theory lives exactly on that tube, which led us to understand the duality as a map between complementary spacetime regions: the three-dimensional world-tube versus the remaining four-dimensional bulk spacetime. Clearly, this is the very definition of a \textit{hologram}. Before defending such claims, though, let us first take the dual theory to the quantum level.\\


\section{Noncommutative dual theory}\label{SubsectionFuzzy}
Thus far, besides the some-what surprising appearance of the self-dual world-tube, we have not really looked into the quantum aspects of the spin-orbit duality. Before doing so, we notice that the hard core of this duality, which is the Hodge $(1+3)$ decomposition w.r.t. a timelike four-momentum, may refer to either a collection of various four-vectors (corresponding to various invariant masses) or just one four-vector. In the quantum level, the former is about the quantum field theory of multi-particle states (i.e. particles of different type) and the latter about the quantum mechanics of single-particle states\footnote{There is also the case when there are multiple four-momenta, all corresponding to the same invariant mass, and, hence, representing a multi-particle system of a single particle-type. We leave this case aside to simplify the conversation, since that does not change the quality of the algebraic arguments that follow.}. Naturally, in order to understand the primary impact of the duality on the quantum regime, we assume a particular four-momentum and single-particle states in a Fock space.

As usual, in the absence of any particular assumption, relativistic quantum mechanics comes with the ordinary Heisenberg algebra\footnote{We set $\hbar=1$ from now on, except at some cases where its presence might be physically meaningful, e.g. when we want to emphasize on the role of the Compton wavelength, $\lambda_C=\hbar/m$.},

\begin{equation}
[\hat{x}^\mu,\hat{p}^\nu]\;=\;i\eta^{\mu\nu}\;,\hspace{2cm}[\hat{x}^\mu,\hat{x}^\nu]\;=\;0\;,\hspace{2cm}[\hat{p}^\mu,\hat{p}^\nu]\;=\;0\;,\label{HeisenbergAlgebra}
\end{equation}\\
which, in our context, is the algebra resulting from canonical quantization of the position and momentum of a massive configuration. Nonetheless, our set-up is based on the $(1+3)$ decomposition w.r.t. four-momentum, where four-position decomposes into its timelike and spacelike parts, $x^\mu=y^\mu+h^\mu$. This results into a decomposition of the Heisenberg algebra as well, the timelike part defined by

\begin{equation}
[\hat{y}^\mu,\hat{p}^\nu]\;=\;i\,\frac{\hat{p}^\mu \hat{p}^\nu}{p^2}\;,\hspace{2cm}[\hat{y}^\mu,\hat{y}^\nu]\;=\;0\;,\hspace{2cm}[\hat{p}^\mu,\hat{p}^\nu]\;=\;0\;,\label{TimelikeAlgebra}
\end{equation}\\
where $p^2=-m^2$ is a Poincar\`e Casimir element that commutes with everything else and, thus, is treated as a c-number. This algebra is a consequence of decomposing the Poisson structure of position and momentum, a natural quantization scheme given in Appendix \ref{AppA}. More importantly, though, since $y^\mu\mapsto y^\mu$ and $p^\mu\mapsto p^\mu$ under the spin-orbit duality, this subalgebra stays the same in the dual theory. Hence, the duality map of four-position (\ref{DualPosition}), $x^\mu\mapsto\tilde{x}^\mu=y^\mu+W^\mu/p^2$, implies that the dual four-position operator $\hat{X}^\mu=\hat{y}^\mu+\hat{W}^\mu/p^2$ satisfies the algebra
\begin{equation}
[\hat{X}^\mu,\hat{X}^\nu]\;=\;-i\,\left(\frac{1}{m^4}\right)\left(\hat{W}^\mu \hat{p}^\nu\,-\,\hat{W}^\nu \hat{p}^\mu\,+\,\epsilon^{\mu\nu\rho\sigma}\hat{W}_\rho \hat{p}_\sigma\right)\;,\hspace{1.6cm}[\hat{X}^\mu,\hat{p}^\nu]\;=\;i\,\frac{\hat{p}^\mu \hat{p}^\nu}{p^2}\;, \label{DualAlgebra}
\end{equation}\\
where we renamed the dual operator $\hat{\tilde{x}}^\mu\equiv\hat{X}^\mu$, to avoid ugly expressions. The rest of the commutators between operators are implied by the usual algebra of the Pauli-Lubanski generator. This algebra is manifestly Lorentz and, also, translation-invariant. Naively, one may think that, since $\hat{X}^\mu=\hat{y}^\mu+\hat{W}^\mu/p^2$, translations in dual space spoil the position commutator but, as illustrated in Appendix \ref{AppendixEMTsusy}, spin-orbit duality preserves translation invariance which strongly suggests that this is not the case. Indeed, $\hat{W}^\mu$ may be the spacelike part of the dual position operator but, at the same time, it is the Pauli-Lubanski operator of the initial theory. In other words, it may pose as dual position but, structurally, it behaves as $\hat{W}^\mu$. This means that, in fact, it does not have a continuous spectrum but eigenvalues accustomed to the particular configuration, it shifts only as a Lorentz (pseudo-)vector (satisfying its defining algebra $[\hat{W}^\mu,\hat{W}^\nu]=-i\epsilon^{\mu\nu\rho\sigma}\hat{W}_\rho\hat{p}_\sigma$) and it does certainly not change as $\hat{W}^\mu\rightarrow\hat{W}^\mu+\hat{W}^\mu_0$ for some operator $\hat{W}^\mu_0\propto\hat{\mathbb{1}}$. The part that does shift as a translation in the dual position operator is $\hat{y}^\mu$, i.e. $\:\hat{y}^\mu\rightarrow\hat{y}^\mu+\hat{y}^\mu_0$ (the same as in the initial theory) but this timelike part is not involved in the r.h.s. of the dual algebra (\ref{DualAlgebra}).

In the rest frame, as seen from (\ref{DualPosition}) (or straight away from (\ref{DualPositionSpatialCOM})), the dual three-position is $\tilde{x}^i=W^i/p^2=mS^i/p^2$, where the associated spin operators $\hat{S}^i$ are the usual generators of the little group satisfying $[\hat{S}^i,\hat{S}^j]=i\epsilon^{ijk}\hat{S}_k$. This implies the subalgebra

\begin{equation}
[\hat{X}_+^i,\hat{X}_+^j]\;=\;\left(\frac{1}{m^2}\right)i\,\epsilon^{ijk}\hat{S}_k\;,\label{DualAlgebraSpatialSpin}
\end{equation}\\
with $\hat{X}^i_+$ being the rest-frame position operators. Equivalently, this subalgebra is a reduction of the Lorentz-invariant one above, (\ref{DualAlgebra}), when the latter is understood to act on states of vanishing three-momentum $\ket{p^i=0,s}$, i.e. $\left\langle\hat{p}^i\right\rangle=0$, which serves as another definition of the rest frame. The mass factor in the r.h.s. of the algebra and the fact that $\hat{X}_+$ is a representation of the spin operator $\hat{S}$ imply a fundamental scale which is defined by the Casimir elements of the underlying spacetime algebra. We illustrate this later on, when we present the induced uncertainty relations.\\

Surprisingly, the expressions we found are not just any kind of noncommutative algebras. The rest-frame spatial algebra (\ref{DualAlgebraSpatialSpin}) is the so-called \textit{spin noncommutativity}, which has been constructed \cite{Falomir:2009cq} and studied from various points of view \cite{Gamboa:2009zs}. Hints that probably inspired such a construction had already been given sometime ago by Jackiw and Nair \cite{Jackiw:2000tz}. Not only that, the generic Lorentz-invariant algebra (\ref{DualAlgebra}) actually matches exactly what was proposed in \cite{Gomes:2010xk} and further studied in \cite{Stechhahn:2019bbe} as a relativistic generalization of the preceding one\footnote{Notice, also, that, in order to match the conventions of \cite{Gomes:2010xk,Stechhahn:2019bbe,Ferrari:2012bv,Lukierski:2013gaa,GuzmanRamirez:2013ynp}, our position commutator in (\ref{DualAlgebra}) may be equivalently expressed as $[\hat{X}^\mu,\hat{X}^\nu]\;=\;i\left(\frac{1}{m^2}\right)\frac{1}{2}\epsilon^{\mu\nu\rho\sigma}\hat{S}_{\rho\sigma}-i\left(\frac{1}{m^4}\right)\epsilon^{\mu\nu\rho\sigma}\hat{W}_\rho \hat{p}_\sigma$, up to numerical prefactors for each of those papers. As we see in the following section, in (\ref{SelfDualOperator}), the particular form presented here is such that the r.h.s. of this algebra conspires into a particular operator of a very special meaning.}, in pursuit of a consistent spin-noncommutative field theory which respects Lorentz symmetry, as opposed to the seminal noncommutative Lorentz-breaking theories \cite{Connes:1997cr}. In that line of work, a noncommutative Dirac equation has been derived \cite{Ferrari:2012bv}, while, also, this noncommutative algebra came up naturally in quantum twistor theory \cite{Lukierski:2013gaa} and in the dynamics of spinning particles on curved spacetime \cite{GuzmanRamirez:2013ynp}.

\subsection*{Noncommutative Wigner 3-spaces}
As far as the second subalgebra in (\ref{DualAlgebra}) is concerned, this is the only difference between the aforementioned investigations on spin noncommutativity and our construction. That is, in their case, the standard Heisenberg commutator $[\hat{x},\hat{p}]=i\eta^{\mu\nu}$ was assumed to be true, whereas, here, the associated commutator $[\hat{X}^\mu,\hat{p}^\nu]=i\hat{p}^\mu \hat{p}^\nu/p^2$ was obtained, not an assumption but merely a result of the $(1+3)$ decomposition. Regardless, one may wonder what is the meaning of this odd-looking commutator. To have a chance in spotting anything familiar we need a comparison with quantum mechanics and, thus, we take the low-energy limit to obtain
\begin{equation}
[\hat{X}^i,\hat{p}^0]\;=\;-i\,\frac{\hat{p}^i}{\hat{p}^0}\;,\label{NWvelocity}
\end{equation}\\
which is the rightful statement that the free particle (i.e. $\dot{p}^\mu=0$) moves at the expected three-velocity (up to orientation, because of the minus sign), given its particular four-momentum. This condition was, actually, introduced by Newton and Wigner \cite{Newton:1949cq} in their effort to acquire a well-defined position operator for massive relativistic particles. Their operator was called the \textit{Newton-Wigner position operator} and its associated scheme the \textit{Newton-Wigner localization} \cite{Jordan:1980er}. The difference between the Newton-Wigner theorem and our results, is that our position operator does not commute with itself, i.e. (\ref{DualAlgebra}), a state of affairs that has actually been suggested as an improvement of this kind of localization \cite{Kasimov}. In any case, as we show in Appendix \ref{AppendixWigner}, the link of our noncommutative structure with the Newton-Wigner operator already takes place in the initial theory and it is, actually, irrespective of the context of the spin-orbit duality.\\

\subsection{Fuzzy de Sitter space}\label{SectionDeSitter}
Even more surprisingly though, in view of the $(1+3)$ decomposition (\ref{BivectorDecomposition}) of bivectors, the position subalgebra in (\ref{DualAlgebra}) actually implies the form

\begin{equation}
[\hat{X}^\mu,\hat{X}^\nu]\;=\;-i\,\left(\frac{1}{m^4}\right)\hat{W}^{\mu\nu}\;,
\end{equation}\\
where $\hat{W}^{\mu\nu}$ is the self-dual total-angular-momentum operator,

\begin{equation}
\hat{W}^{\mu\nu}\;\equiv\;\hat{W}^\mu \hat{p}^\nu\,-\,\hat{W}^\nu \hat{p}^\mu\,+\,\epsilon^{\mu\nu\rho\sigma}\hat{W}_\rho \hat{p}_\sigma\;=\;\hat{M}^{\mu\nu}|_{\mbox{\tiny E=H}}\;.\label{SelfDualOperator}
\end{equation}
This is the operator associated with a total angular momentum $M^{\mu\nu}$ which is self-dual, i.e. $M^{\mu\nu}=\star M^{\mu\nu}$, or, in other words, an angular momentum that has equal electric and magnetic parts, i.e. $E^\mu=H^\mu=W^\mu$. Of course, being an angular-momentum operator, $\hat{W}^{\mu\nu}$ naturally satisfies the $\mathfrak{so}(1,3)$ Lorentz algebra.\\

For one, this form of the position algebra resembles most of the covariant noncommutative structures in the literature \cite{Snyder} and most notably \cite{Much:2017hcv}. More importantly, though, the fact that this position algebra projects onto a SO$(1,3)$ generator implies that the underlying space may be equally seen as a \textit{three-dimensional noncommutative de Sitter space}. Whereas, indeed, in the previous section we showed that the (classical) dual theory lives in a three-dimensional de Sitter world-tube. Fuzzy de Sitter space has been studied before, but usually in two and four dimensions \cite{Gazeau:2006hj}. In any case, it is noteworthy that, except being self-dual, this operator squares to zero identically,

\begin{equation}
\hat{W}^{\mu\nu}\;=\;\star\hat{W}^{\mu\nu}\hspace{2cm}\mbox{and}\hspace{2cm}\hat{W}^{\mu\nu}\hat{W}_{\mu\nu}\;=\;0\;.
\end{equation}
Those are statements in the dual theory, where spin and position exchange roles: what was before realized as a (particular) spin vector, in this picture is a spacelike position $\tilde{h}^\mu$ living on a (particular) de Sitter world-tube. Hence, self-duality selects a dual spin vector $\tilde{W}^\mu$ such that $\tilde{W}^\mu/p^2=\tilde{h}^\mu$, where $\tilde{h}^\mu=W^\mu/p^2$. The operator $\hat{W}_{\mu\nu}$ projects on such dual states and, since $\hat{W}_{\mu\nu}^2=0$, those are of a vanishing total angular momentum.\\

Before digging any deeper, we feel we should pause to appreciate all this. In our understanding, this is the second striking result of the spin-orbit duality: the dual quantum theory is noncommutative, a fact that came up in a completely natural way. The only caveat one could impose against the universality of this statement is that, from the beginning, we have assumed a first-order action principle, $S=S[q,\partial q]$. But that should not pose a problem of any sort since, first of all, the vast majority of the known field theories are first-order, avoiding the usual pathology of vacuum states whose energy is unbounded from below. Even for higher-derivative theories, though, currents would just be supplemented with additional terms, not altering the actual conservation laws and the structural interplay between spin and orbital angular momentum, showing, therefore, no indication whatsoever that the spin-orbit duality would seize to exist. It would be instructive to prove this rigorously.\\

One could also wonder whether us studying single-particle states makes noncommutativity dedicated solely to that particular case. But this too is not true. The reason we chose single-particle states is really because, then, we are dealing with quantum mechanics where a position operator is a well-defined object in the underlying algebra, a state of affairs that is simply not there for the quantum field theory of multi-particle states. Of course, since quantum mechanics is noncommutative, the same must hold for the associated quantum field theory. In each case, the noncommutative scale is set by the Casimir element $p^2=-m^2$.\\

At this point, we should note that we have not yet mentioned anything about the usual Lagrange equations of motion. We intentionally bring this up now that we have a more complete view of the dual theory. For one, four-momentum is invariant which implies that momentum space stays the same. In that respect, invariance of the equations of motion would imply that this duality is also a true symmetry of massive Poincar\`e or Lorentz-invariant theories. On the other hand, considering position space, we fail to see how a map between a commutative and a noncommutative structure can be an honest symmetry. In any case, for now, we hold on to the fact that spin-orbit duality maps the four-dimensional bulk spacetime onto a noncommutative, three-dimensional world-tube around the center of mass.

\subsection{Fundamental scale and cutoff}\label{SectionPoleMass}
The mass which characterizes the noncommutative algebra is a Poincar\`e Casimir element and, therefore, should always account for the physical, pole mass $m_P$. Hence, all this time we really meant $p^2=-m^2=-m_P^2$. This is the mass corresponding to external states, identified with the pole in propagators. Naturally, for a free theory this equals the bare mass $m_0$ in the (bare) Lagrangian, $m_P=m_0$. In a fully-interacting theory, the mass gets corrected and renormalized, while, depending on the subtraction scheme, the renormalized mass, $m_R$, may or may not coincide with the pole mass. Still, whatever the case might be, the Casimir element $-$and, thus, the fundamental scale$-$ is always identified with the pole mass and is, therefore, independent of the energy scale\footnote{This makes sense, since, if the opposite was true, such a scale-dependent noncommutative algebra would imply, as energy grows, an ever-shrinking fundamental length. In turn, this would also imply an ever-growing UV cutoff which, by its definition and purpose, makes no sense either.}. Hence, this is tantamount to setting the UV cutoff equal to $m_P$,
\begin{equation}
\Lambda_{\mbox{\tiny UV}}\;\sim\;m_P
\end{equation}\\
in the dual noncommutative theory. At the same time, a finite noncommutative space implies also an infrared (IR) cutoff, of the order of its radius. Hence, since both cutoffs depend on the mass, there is a direct tension between them, which is a non-perturbative manifestation of the UV-IR mixing phenomenon \cite{Minwalla:1999px,Horvat:2011bs}. Anyhow, it is intriguing to investigate the impact of this peculiar cutoff on the renormalization group, since it is at $\lambda_C=\hbar/m$ where quantum effects begin to become important.

\subsection{Fuzzy sphere}\label{SubsectionFuzzySphere}
Interestingly, there is another way to view the position commutator in (\ref{DualAlgebra}). That is, the commutator is equivalent to

\begin{equation}
[\hat{X}^\mu,\hat{X}^\nu]\;=\;-i\,\left(\frac{1}{m^2}\right)\left(\hat{X}^\mu \hat{p}^\nu\,-\,\hat{X}^\nu \hat{p}^\mu\,+\,\epsilon^{\mu\nu\rho\sigma}\hat{X}_\rho \hat{p}_\sigma\right)\;,
\end{equation}\\
where, as it is easy to see, if we expand $\hat{X}^\mu=\hat{y}^\mu+\hat{W}^\mu/p^2$, and given the definition $\hat{y}^\mu=(\hat{x}\cdot\hat{p})\hat{p}^\mu/p^2$, all terms involving $\hat{y}^\mu$ cancel or vanish identically and we get back to the original form (\ref{DualAlgebra}). Of course, as explained below (\ref{DualAlgebra}), since those $\hat{y}$-terms do not really contribute, despite appearances, the above commutator is (still) translationally invariant. The reason we expressed it that way is that, now, a Lie-algebraic form is manifest. In fact, by removing the third term in the r.h.s. of the above expression and taking four-momentum as a constant vector, then the remaining algebra is associated with the so-called \textit{$\kappa$ deformations} of the Poincar\`e Hopf algebra \cite{Lukierski}; those deformations did not originally preserved the Lorentz algebra, as opposed to subsequent constructions \cite{Kehagias:1993ju}. When considered in the rest frame, i.e. when it acts on states $\ket{p^i=0,s}$ such that $\hat{p}^i=0$ and $\hat{p}^0=m$ can be understood, the algebra effectively reduces to

\begin{equation}
[\hat{X}^i_+,\hat{X}^0_+]\;=\;-i\,\left(\frac{1}{m}\right)\hat{X}^i_+\;,\hspace{2cm}[\hat{X}^i_+,\hat{X}_+^j]\;=\;-i\,\left(\frac{1}{m}\right){\epsilon^{ij}}_k\hat{X}_+^k\;.\label{kMinkowski}
\end{equation}\\
Again, if $[\hat{X}^i_+,\hat{X}_+^j]=0$ was true, then we would have the so-called \textit{$\kappa$-Minkowski spacetime} \cite{Daszkiewicz:2007eq} where the fundamental scale would be identified with $\kappa=m$. Nonetheless, the reason for reducing to the rest-frame is that the second commutator above yields a very special kind of noncommutative space.\\

As a matter of fact, there is an equivalent way to acquire this algebra. Taking into account, one more time, that the dual rest-frame position operator is $\hat{X}_+^i=-\hat{W}^i/m^2=-\hat{S}^i/m$, we realize that spin-noncommutativity (\ref{DualAlgebraSpatialSpin}) is, equivalently,

\begin{equation}
[\hat{X}_+^i,\hat{X}_+^j]\;=\;i\,\lambda_C\,{\epsilon^{ij}}_k\,\hat{X}_+^k\;,\label{FuzzySphere}
\end{equation}\\
where we absorbed an orientation-related minus sign of the r.h.s. into the definition of the Levi-Civita symbol. Naively, this algebra seems to break translation invariance, contradicting the translationally invariant generic algebra (\ref{DualAlgebra}) and, equally, as restated, the fact that spin-orbit duality preserves translation invariance. But, as both of those arguments strongly suggest, no such breaking occurs. That is, as was illustrated in and around (\ref{DualSphere}), the dual theory, in the rest frame, sits on a fixed two-sphere of the M\o{}ller radius $\rho$, which means that rest-frame translations in the dual theory are actually SU$(2)$ rotations on a two-sphere. And so (\ref{FuzzySphere}), which is an enveloping algebra of $\mathfrak{su}(2)$, is invariant under the action of its own elements. In other words, since $\hat{X}_+^i=-\hat{S}^i/m$ in the dual picture, $[\hat{X}_+^i,\hat{X}_+^j]=i\lambda_C{\epsilon^{ij}}_k\hat{X}^k_+$ is preserved under translations, in the exact same manner that the little-group algebra $[\hat{S}^i,\hat{S}^j]=i\epsilon^{ijk}\hat{S}_k$ is preserved under rotations. Notice that this invariance is a property in the particular context of the spin-orbit duality, where dual position realizes (i.e. it is a label for) the spin vector of the initial theory. For a general Lie-algebraic position algebra this, of course, is not true. \\

Given the exact structure of (\ref{FuzzySphere}), this is not just any position-dependent, Lie-algebraic noncommutative algebra either. It is a fuzzy sphere \cite{Madore:1991bw} of radius

\begin{equation}
R_\circledast\;\equiv\;\sqrt{\hat{X}_+^2}\;=\;\sqrt{\frac{\hat{S}^2}{m^2}}\;=\;\rho\;=\;\lambda_C\,\sqrt{s(s+1)}\;,\label{FuzzySphereRadius}
\end{equation}\\
which is, as expected from (\ref{DualSphere}), the M\o{}ller radius and, as usual, $\hat{S}^i$ are understood as $(2s+1)$-dimensional operators acting on a space of massive spin states, $\lbrace\ket{s,m_s,m}\rbrace_{m_s=-s}^{m_s=+s}$. Then, the algebra (\ref{FuzzySphere}) may, as well, be written as

\begin{equation}
[\hat{X}_+^i,\hat{X}_+^j]\;=\;\left(\frac{R_\circledast}{\sqrt{s(s+1)}}\right)i\,{\epsilon^{ij}}_k\,\hat{X}_+^k\;,\label{FuzzySphere2}
\end{equation}\\
which is a universal enveloping algebra U$\left(\mathfrak{su}(2)\right)$ of the $\mathfrak{su}(2)$ algebra of the spin operators, $[\hat{S}^i,\hat{S}^j]=i\epsilon^{ijk}\hat{S}_k$, for
\begin{equation}
\hat{X}_+^i\;=\;-\frac{1}{m}\;\hat{S}^i\;=\;-\frac{R_\circledast}{\sqrt{s(s+1)}}\;\hat{S}^i\;.
\end{equation}\\
Hence, the dual space is discrete, in the sense that the spectrum of eigenvalues of the position operators $\hat{X}^i$ is of dimension $2s+1$.

\subsubsection*{Dual dispersion rings}
In fact, the fact that $\hat{X}_+^i=-\hat{S}^i/m$, or, in other words, that the two noncommutative algebras

\begin{equation}
[\hat{X}_+^i,\hat{X}_+^j]\;=\;i\,\lambda_C\,{\epsilon^{ij}}_k\,\hat{X}_+^k\hspace{1cm}\Leftrightarrow\hspace{1cm}[\hat{X}_+^i,\hat{X}_+^j]\;=\;\left(\frac{1}{m^2}\right)i\,\epsilon^{ijk}\hat{S}_k\;,\label{FuzzyVSspin}
\end{equation}\\
are equivalent, makes this fuzzy sphere of a quite special kind. First, we understand that, in this case, a mean position on the sphere makes sense, since

\begin{equation}
\langle\hat{X}_+^i\rangle\;=\;-\frac{\langle\hat{S}^i\rangle}{m}\;,
\end{equation}\\
such that a configuration in some spin state has a particular expected localization on it. Hence, except the inherent scale given by the mass of the configuration, there is, at the same time, a built-in orientation in the dual theory. This localization, though, can be made more exact. That is, the spin-noncommutative algebra, second of the two in (\ref{FuzzyVSspin}), implies an uncertainty relation

\begin{equation}
\Delta\hat{X}_+^i\,\Delta\hat{X}_+^j\;\geq\;\frac{1}{m^2}\,\abs{\left\langle\epsilon^{ijk}\hat{S}_k\right\rangle}\;,\label{DispersionSpin}
\end{equation}\\
where, for example, a state of maximal spin-projection along the $Z$-dimension has

\begin{equation}
\Delta X\,\Delta Y\;\geq\;\frac{s}{m^2}\;,\label{XYdispersion}
\end{equation}\\
with $s$ the spin of the field, while it holds that $\,\langle\hat{Z}\rangle=-s/m\,$ and $\,\langle\hat{X}\rangle=\langle\hat{Y}\rangle=0$. In fact, it may be that the mean values of $\hat{X}$ and $\hat{Y}$ vanish, but since their dispersion is nonzero and satisfies (\ref{XYdispersion}), then this inequality defines a (minimum) radius in the $X-Y$ plane, i.e. $r_{\mbox{\tiny XY}}\geq\sqrt{s}/m$. In order to understand what this means, it instructive to become a bit more specific and consider a spinor field representation of spin $s=1/2$.\\

\begin{figure}[t!]
    \centering
    {{\includegraphics[width=14cm]{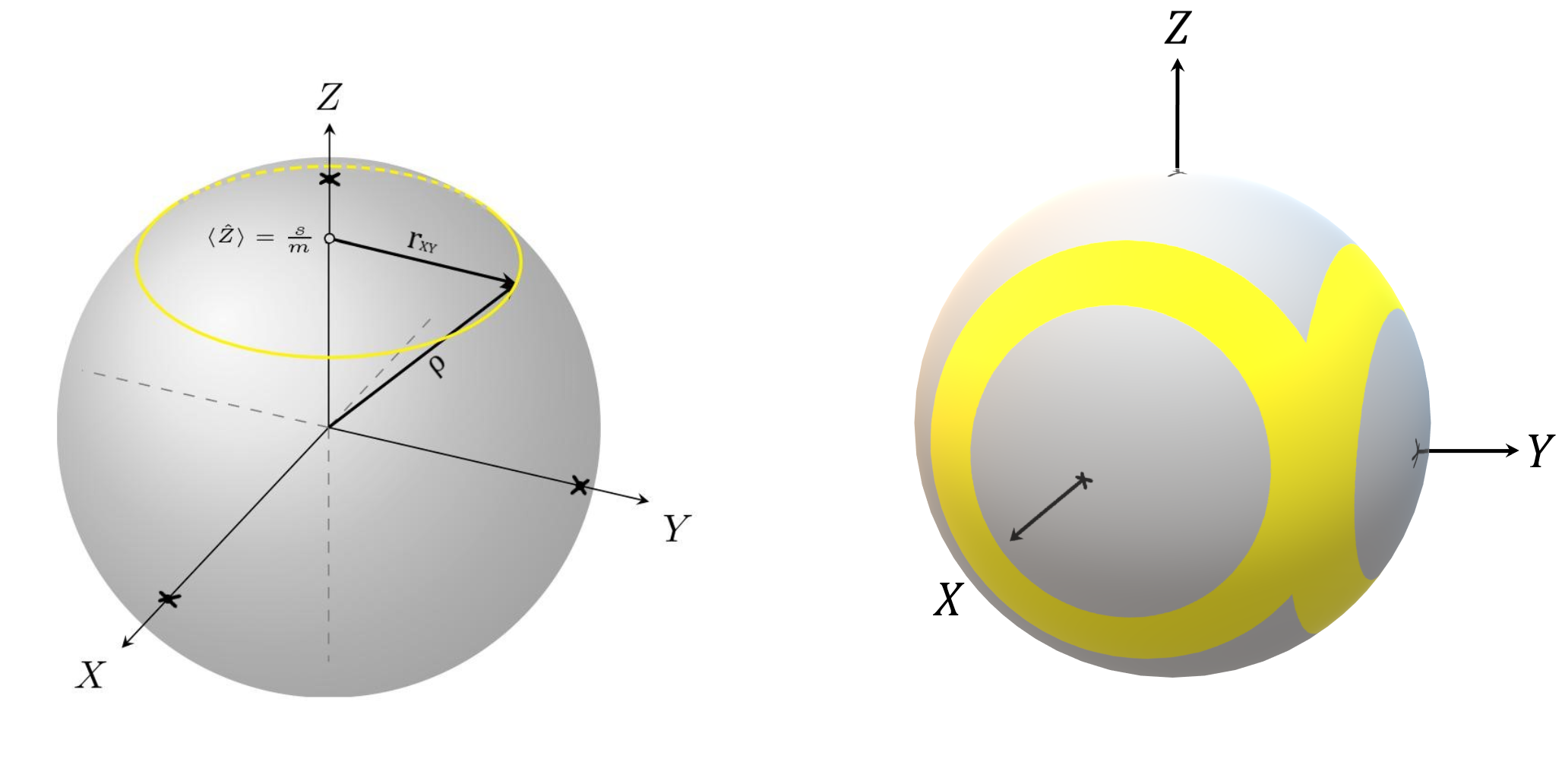} }}%
\caption{Representations of two $s=\frac{1}{2}$ states in the dual fuzzy sphere of the M\o{}ller radius. a)On the left: the dual picture of a maximal-spin spinor-state with $\,\langle\hat{S_z}\rangle=-m\langle\hat{Z}\rangle=s\,$ and, thus, $\,\langle\hat{S_x}\rangle=\langle\hat{X}\rangle=\langle\hat{S}_y\rangle=\langle\hat{Y}\rangle=0$. This is a (yellow) ring around the $Z$ axis, on which the field excitation in dispersed, according to relations $\Delta X\Delta Y\geq r_{\mbox{\tiny XY}}^2$ and $\Delta Z=0$. The ring is localized along the $Z$ dimension on $\langle\hat{Z}\rangle=\frac{s}{m}$, since $\Delta Z=0$, where we enforced a positive sign just as a matter of taste; as stated, spin-orbit duality is true up to a sign. b)On the right: a superposition state with $\langle\hat{S_x}\rangle=\langle\hat{S}_y\rangle=\frac{s}{\sqrt{2}}$ and, thus, $\,\langle\hat{S_z}\rangle=0$. The example used is $\ket{s}=\left(\frac{1+i}{2}\right)\ket{+}-\left(\frac{i}{\sqrt{2}}\right)\ket{-}$, where $\ket{\pm}$ are the basis vectors of spin projection along the $Z$. Ribbons have replaced rings, since now $\Delta X$, $\Delta Y$, $\Delta Z\neq0$.}
\label{rings}
\end{figure}

For the $\,s=1/2\,$ field, $\,\Delta S_i=\langle\hat{S}_i^2\rangle-\langle\hat{S}_i\rangle^2=s^2-\langle\hat{S}_i\rangle^2\,$ and, thus, $\,\langle\hat{S_z}\rangle=-m\langle\hat{Z}\rangle=s\,$ implies $\Delta S_z=\,\Delta Z=0\,$, which exactly localizes the configuration in the $Z$-dimension\footnote{Of course, we should not be worried about the statement that, given $\,\Delta Z=0\,$, the configuration is localized along the $Z$-dimension, since $\,\Delta X\,$ and $\,\Delta Y\,$ are still nonzero. In other words, the configuration \textit{is} delocalized, with all its uncertainty dispersed along a $X-Y$ ring on the fuzzy sphere. This is depicted in Figure \ref{rings}a. Moreover, one could argue that, since $\,\langle\hat{S_x}\rangle=-m\langle\hat{X}\rangle=\langle\hat{S}_y\rangle=-m\langle\hat{Y}\rangle=0$ imply $\Delta S_x=m\Delta X=\Delta S_y=m\Delta Y=s$, the naive product $\Delta X\Delta Y=s^2/m^2$ contradicts the uncertainty relation (\ref{XYdispersion}), $\Delta X\,\Delta Y\;\geq\;\frac{s}{m^2}$. But, as exactly in standard quantum mechanics, this is not the case: an uncertainty product is given as an inequality with regard to the underlying algebra, e.g. (\ref{DispersionSpin}), not by multiplying individual dispersions.}. Therefore, because the configuration lives on the fuzzy sphere, the radius $r_{\mbox{\tiny XY}}$ dictated by the uncertainty relation (\ref{XYdispersion}) defines a ring on that sphere, around the $Z$-axis. And since this sphere is of the M\o{}ller radius, it is straightforward to check that the radius of the ring must be
\begin{equation}
r_{\mbox{\tiny XY}}^2\,+\,\langle\hat{Z}\rangle^2\;=\;\rho^2\;=\;\frac{s(s+1)}{m^2}\hspace{1cm}\Rightarrow\hspace{1cm}r_{\mbox{\tiny XY}}^2\;\equiv\;(r_{\mbox{\tiny XY}}^{\mbox{\tiny min}})^2\;=\;\frac{s}{m^2}\;,
\end{equation}\\
which is, indeed, the minimum value permitted by (\ref{XYdispersion}). This is all depicted in Figure \ref{rings}a. Overall, in the rest frame of the dual theory, such a state of a field configuration is understood to be a dispersed excitation along a ring on a fuzzy sphere. Conversely, such a \textit{dispersion ring} around the $Z$-axis in the dual sphere, corresponds to a state of $\,\langle\hat{S_z}\rangle=s\,$ in the initial theory.\\

Needless to say, analogous dispersion rings are dual to maximal-spin states in the $Y$ or $X$ dimensions too. However, when superposition of states is considered, a somewhat different picture arises, where rings are replaced with ribbons, as schematically visualized in Figure \ref{rings}b. Nonetheless, the analogy remains the same: states with spin projections in one axis in the initial theory, correspond to ribbons around the same axis in the dual fuzzy sphere.

\section{A holographic interpretation}\label{SubsectionHoloInter}
Nevertheless, exchanging between spatial and spin degrees of freedom still sounds somewhat extraordinary if not absurd. What does it really mean to have a dual theory where what was before realized as an internal degree of freedom is now an external one? To answer such questions, we now attempt a full-scale interpretation of the spin-orbit duality.\\

Let us, first, recap and see the big picture here. To keep things simple, we go to the rest frame, where we understand that the duality acts on a restricted range of the three-position of the center of mass,
\begin{equation}
\abs{\vec{x}}\;>\;\rho\;=\;\frac{S}{m}\;.
\end{equation}\\
This restriction is because the spin-orbit duality naturally selects the surface of a two-sphere of the M\o{}ller radius $\rho$ as its self-dual region, on which the duality becomes the trivial map. Incidentally, as explained, relativity implies non-covariance on and inside the ball of radius $\rho$, while quantum mechanics poses the same limitation since at $\hat{\rho}\approx\lambda_C$ pair-production comes up and the concept of position breaks down anyway. In the dual theory, on the other hand, the rest-frame three-position is

\begin{equation}
\tilde{x}^i\;=\;-\frac{S^i}{m}\hspace{1cm}\Rightarrow\hspace{1cm}\tilde{x}^i\tilde{x}_i\;=\;\rho^2\;=\;\frac{S^2}{m^2}\;,
\end{equation}\\
which says that the configuration lives exactly on that two-sphere of the M\o{}ller radius $\rho$. Therefore, we deduce that all three-position of the initial theory, except an open ball of radius $\rho$, is mapped into the boundary two-sphere of that very ball. In simple words, in the rest frame, the duality is a map between complementary spacetime regions: the two-dimensional sphere versus the three-dimensional bulk spacetime. What is realized as spin in the bulk is orbital angular momentum on the two-sphere and vice versa. In particular,

\begin{equation}
L_{\mu\nu}\hspace{1cm}\mapsto\hspace{1cm}\tilde{L}_{\mu\nu}=\star S_{\mu\nu}\;,
\end{equation}\\
says, for example, that the (dual) $z$-position on the two-sphere is given by the (initial) projection of spin in the bulk, along the same dimension,

\begin{equation}
\tilde{z}\;=\;-\frac{S_z}{m}\;.
\end{equation}\\
The significance of this statement, though, is better understood in quantum mechanics. That is, as illustrated in the last section, the associated state is defined by

\begin{equation}
\langle\hat{Z}\rangle\;=\;-\frac{\langle\hat{S}_z\rangle}{m}\hspace{2cm}\mbox{and}\hspace{2cm}\Delta X\Delta Y\;=\;r_{\mbox{\tiny XY}}^2\;=\;\frac{\langle\hat{S}_z\rangle}{m^2}\;.
\end{equation}\\
This reflect a dispersion ring of radius $r_{\mbox{\tiny XY}}$, around the $z$ axis, on the dual fuzzy sphere, as depicted in Figure \ref{rings}a. But, in turn, such a dispersion ring implies orbital angular momentum on the $XY$ plane, or, in other words, a $\Delta L_z$ component on the fuzzy sphere; this can be understood as the fact that the dispersion product $\Delta X\Delta Y$ implies the existence of an uncertainty $\Delta L_z\Delta\phi$, where $\Delta\phi$ is the angular position about the $z$ axis, on the $XY$ plane. Let us rephrase: orbital angular momentum along an axis in the two-sphere represents spin angular momentum, along the same axis, in the remaining bulk three-space. Eventually, this swapping between kinematic and spin degrees of freedom is the very incarnation of the spin-orbit duality.\\

So let us generalize all this for any frame and make sense of the situation. The duality is eventually a map between complementary regions of spacetime. That is, between a three-dimensional word-tube with radius of the order of the Compton wavelength $\lambda_C$ $-$around the center of mass of the configuration$-$ and the remaining four-dimensional bulk spacetime. Spin and orbit degrees of freedom exchange roles and what was before realized as spin is now orbital angular momentum, i.e. $\tilde{L}_{\mu\nu}=\star S_{\mu\nu}$, which makes an exact analogy with what was discussed in the end of Section \ref{SectionSpin} where spin emerged as bound currents of angular momentum. As a matter of fact, in the quantum level, actual orbital-angular-momentum dispersion rings in the (dual) noncommutative world-tube correspond to spin of quantum states in the bulk (initial) theory. Needless to say, all four-dimensional information being stored onto a three-dimensional de Sitter surface (or a flat cylinder, in the rest frame) is the very definition of a hologram. This state of affairs naturally drive us to interpret spin-orbit duality as a \textit{holographic} map. External and internal degrees of freedom in the bulk Minkowski space outside the world-tube $-$the configuration properties that define scales larger than $\lambda_C-$ are inversely encoded onto this natural boundary. And vice versa.\\

From this point of view, an exchange between internal and external degrees of freedom when we shift between the dual pictures $-$the one on the tube versus its higher-dimensional projection$-$ makes total sense. Spatial (orbit) degrees of freedom in the bulk space (initial theory) could never be mapped into spatial degrees of freedom on a lower-dimensional surface, much like points in a vector space cannot be stereographically projected down onto a lower-dimensional subspace without overlapping. Whereas, indeed, spin-orbit duality indicates that spatial degrees of freedom are encoded as spin on the dual world-tube, $\tilde{S}_{\mu\nu}=\star L_{\mu\nu}$. In reverse, orbital angular momentum on the dual world-tube manifests as spin degrees of freedom in the higher-dimensional bulk space, $\tilde{L}_{\mu\nu}=\star S_{\mu\nu}$, since kinematics on a compact surface may only provide bound currents of orbital angular momentum w.r.t. the higher-dimensional space in which the surface is embedded. Again, those are the bound currents of the improved energy-momentum current $\Theta^{\mu\nu}$ (which incorporates spin) in four-dimensional Minkowski space. Hence, it looks like we are dealing with a legitimate hologram and spin-orbit duality should be probably understood as a realization of the holographic principle \cite{Stephens:1993an}.\\

Therefore, this duality may as well be seen as a link between large and small scales, i.e. the bulk spacetime and a de Sitter world-tube of the Compton scale. In turn, since this world-tube is shown to be noncommutative, this naturally suggests that, for a Poincar\`e or just Lorentz-invariant quantum theory with massive fields of nonzero spin, spacetime is quantized at the fundamental level.


\section{Summary and directions}\label{SectionSummary}
Summarizing, a new duality was proposed in four-dimensional flat space, which exchanges between spin and orbital degrees of freedom. This was motivated by a Hodge decomposition of the angular-momentum bivector for massive fields, along which spin and orbital angular momentum are Hodge duals of one another. The duality respects Poincar\`e symmetry and was shown to transform between complementary spacelike regions, projecting a fixed three-dimensional de Sitter world-tube of the Compton scale (around the center of mass) into the bulk of four-dimensional spacetime and vice versa. This was interpreted as a realization of the holographic principle. Surprisingly, the dual theory living on that tube turns out to be noncommutative and entirely defined by the Casimir elements of the Poincar\`e algebra. In fact, the pole mass is the ultraviolet cutoff. Ultimately, this naturally suggests that, for a Poincar\`e or just Lorentz-invariant quantum theory with massive fields of nonzero spin, spacetime is quantized at the fundamental level.\\

Before discussing any possible directions, we shall underline a couple of facts. The first, obviously, is the suggestion that a large class of field theories exhibit a noncommutative spacetime. This is, of course, profound because it implies that any such theory could, in principle, be UV finite. The second point which we feel deserves attention is the fact the central quantities predicted by the duality $-$the world-tube radius and the noncommutative algebra$-$ are identified with objects already present in the literature. Those would be the M\o{}ller radius and spin/de Sitter noncommutativity. For one, the M\o{}ller radius roots in both relativity and quantum mechanics and poses a limitation in localization, which is perfectly in line with the fact that spin-orbit duality becomes trivial on this exact region. Secondly, while spin/de Sitter noncommutativity are arbitrary constructions in the literature which serve as consistent algebras that respect Lorentz symmetry (as opposed to the seminal noncommutative Lorentz-breaking theories), here they are derived as a natural feature of all massive field theories with spin, without any sort of assumptions. It is truly remarkable the way spin-orbit duality provides an arena where all those concepts and noncommutative algebras meet.

\subsection*{Discussion and ideas}
So, first of all, let us make a comment on the bold statement that spacetime is quantized. It is clear that spin-orbit duality, among other points of view, serves as a link between the large and small scales, implying that massive fields with spin exhibit a noncommutative geometry at the Compton scale. Nonetheless, it is not clear how to implement this in a field theory. Should we simply choose an appropriate quantum field theory and consider it in a spacetime quantized according to spin-noncommutativity? Or, maybe, are there other objects that transform along, according to the duality map? If the first case were true, we would just employ the spin-noncommutative algebra for a massive scalar or Dirac field in the presence of a U$(1)$ gauge field. Or the Proca theory. Or, even better, the Abelian Higgs model \cite{Lozano:2000qf}. Then, the Moyal product \cite{Moyal:1949sk} would be enforced and propagators would be calculated. In that case, another privilege of this duality would be the offer of the pole mass as a UV cutoff. In any case, the exact application of the spin-orbit duality on noncommutative field theory merits further investigation.\\

Next, we strongly believe that there should be some sort of connection between the spin-orbit duality and S-duality \cite{Sen:1994fa}. This is mostly because of the obvious structural similarity between the spin-orbit and the electromagnetic duality, the latter being a simple case of the Montonen-Olive duality \cite{Montonen:1977sn} of gauge theory, which in turn is an example of S-duality. Another hint, in this direction, is the tension that has been observed between the electromagnetic duality and the separation between spin and orbital angular momentum \cite{Bliokh:2012zr}. The simplest investigation to that end should probably initiate by establishing a relation with the standard electromagnetic duality of pure U$(1)$ gauge theory. Such an attempt, however, requires us generalizing the spin-orbit duality for the case of massless fields.\\

In turn, employing the duality for the massless case is not, in any way, guaranteed to work. Of course, if indeed possible, that would let us consider the full conformal group and investigate whether the duality holds for the larger class of conformal field theories (CFTs). In such an occasion, the presence of an intrinsic (noncommutative) scale would maybe look incompatible with conformal symmetry, but this most-probably is a matter of the way symmetry is implemented in quantum geometry \cite{Hatzinikitas:2001ht}. In any case, such a generalization seems impossible even from its very first step, since the projection tensor $h^{\mu\nu}(p)=\eta^{\mu\nu}-p^\mu p^\nu/p^2$ of the $(1+3)$ decomposition diverges when we take the massless limit $m\rightarrow0$. However, we believe that this limiting procedure is naive, as it is usually the case with similar occasions in field theory, while probably the proper way to go would be to shift to lightcone coordinates.\\

Speaking of generalizing the duality for larger symmetry groups, another important matter we have not yet addressed is supersymmetry. Both the Poincar\`e superalgebra and Noether's supercurrent conservation law seem to be invariant under the spin-orbit duality, statements for which a proof is sketched in Appendix \ref{AppendixSUSY}. Preservation of supersymmetry would not only generalize the class of the dual noncommutative spaces but, more importantly, it plays a decisive role in the dynamics and stability of noncommutative field theory \cite{Iso:2001mg}. Nevertheless, a more-formal proof in Minkowski superspace should probably be pursued, an analysis left for future investigation.\\

Anyhow, we suspect that the most interesting viewing angle for this duality is string theory, where we would have to work with reductions of the superstring in four dimensions. The most prominent example would probably be the quantization of D3-branes, although compactifications and more-complex Hannany-Witten D-brane set-ups \cite{Hanany:1996ie} could also be considered. In any case, string theory seems an appealing context for the duality due to various reasons. One of them, is that, in a way, it involves the M\o{}ller radius operator $\hat{\rho}=\sqrt{s(s+1)}\hbar/m$ in the disguise of the Bogomol'nyi-Prasad-Sommerfield (BPS) bound, $s\leq m$. This yields $\hat{\rho}^2\leq(s/m^2+1)\hbar$, which should be somehow related to the fundamental string scale and, maybe, correlated to string-theory uncertainty relations. Another obvious reason is the quantum geometry induced by open strings on D-branes. Moreover, the fact that the dual rest frame always hosts a fuzzy sphere provokes an association with D0-branes and Matrix theory. Nevertheless, we do not rush into conclusions, since those theories involve matrices of dimension $N$, where $N$ the number of D-branes, whereas in the context of the spin-orbit duality the dimension of the (rest-frame) dual-position operator, $\hat{X}^i=-\hat{S}^i/m$, is $\;2s+1$.\\

Eventually, since similarities between the two theories are tantalizing, almost everything we said about string theory may be argued in favor of black holes too. In particular, the analogue of $S/m$ is found in any rotating solution, e.g. the Kerr metric. It would be interesting to find a meeting point between the spin-orbit duality and noncommutative Kerr black holes \cite{DiGrezia:2009di}. In fact, a more general effort to correlate this duality with noncommutative theories probably points to the most prominent non-local concept, i.e. entanglement entropy, which has been studied before in noncommutative space \cite{Dou:2006ni}.\\

Surprisingly, as far as entanglement is concerned, a duality has been shown to exist for entangled systems of identical particles under the exchange between internal and external degrees of freedom \cite{Bose}. This duality was argued to enable a reliable test of quantum indistinguishability \cite{Moreva}, while it was also specified for spin and orbital degrees of freedom \cite{Bhatti} (although applied to (massless) photons, which elude our analysis as of yet). Soon after, this so-called \textit{``entanglement duality''} was generalized for distinguishable particles too \cite{Karczewski}. Anyhow, we suspect that this is a direct realization of the spin-orbit duality presented here.\\

Finally, another practical direction that would shed light onto the swapping between spin and orbit degrees of freedom would be to look what happens to the eigenvalue problem of the angular-momentum operators. That is, considering vortices (rotation eigenmodes) and boost eigenmodes \cite{Biernat:2007sz}, we may wonder what is the physical picture of them exchanging roles under the spin-orbit duality.

\paragraph*{Acknowledgments}
I thank M. Tsamparlis, S. Speziali, G. Savvidy for our useful discussions, C. Nunez and G. Itsios for their comments, V. P. Nair for his suggestions and an anonymous reviewer of an earlier version of the manuscript for pointing out some misunderstandings on noncommutative geometry. I am always grateful to Fani Christidi for our inspiring dialogue.


\appendix
\section{Energy-momentum current from Noether's second theorem}\label{AppendixNoether}
As always, we may prefer a more geometric angle and view Lorentz transformations for what they really are,

\begin{equation}
x^\mu\;\rightarrow\; x'^\mu\:=\:x^\mu+{\omega^\mu}_\nu\, x^\nu\:=\:x^\mu+\xi^\mu(x)\;,\hspace{1cm}\xi^\mu(x)\equiv{\omega^\mu}_\nu\, x^\nu\;,\label{LocalTranslations}
\end{equation}\\
i.e. spacetime-dependent translations. In fact, along this (Killing) vector field, general fields' transformations are naturally identified with its Lie derivative, $\delta q=\frac{i}{2}\omega^{\mu\nu}\mathbf{M}_{\mu\nu}\,q=\mathfrak{L}_\xi q$, and thus the Lagrangian density varies, on the equations of motion, as

\begin{equation}
\partial_\mu\left(\frac{\partial\mathcal{L}}{\partial\partial_\mu q}\mathfrak{L}_\xi q\right)\;=\;\delta\mathcal{L}\;=\;\mathfrak{L}_\xi\mathcal{L}\;=\;\partial_\mu\left(\xi^\mu\mathcal{L}\right)\;,
\end{equation}\\
where, in the last equation, we used that $\partial_\mu\xi^\mu=0$. This implies

\begin{equation}
0=\int_{\mathbb{R}^4}\partial_\mu\left(\frac{\partial\mathcal{L}}{\partial\partial_\mu q}\mathfrak{L}_\xi q-\xi^\mu\mathcal{L}\right)=\int_{\mathbb{R}^4}\partial_\mu\left(\xi_\rho\,\mathcal{T}^{\mu\rho}+\partial_\nu\xi_\rho\mathcal{S}^{\mu\nu\rho}\right)
=\int_{\mathbb{R}^4}(\partial_\mu\xi_\rho)\left(\mathcal{T}^{\mu\rho}+\partial_\nu\mathcal{S}^{\mu\nu\rho}+\partial_\nu U^{\nu\mu\rho}\right)\;,\label{Noether2nd}
\end{equation}\\
where we used the conservation law $\partial_\mu\mathcal{T}^{\mu\rho}=0$, by Noether's first theorem, set $\partial_\mu\partial_\nu\xi_\rho=0$ for Lorentz transformations and integrated by parts two times in the last equation; the total derivative $\partial_\nu U^{\nu\mu\rho}$ reflects the ambiguity of this integration. Since $\partial_\mu\xi_\rho=\omega_{\mu\rho}$, the antisymmetric part of the expression in the parenthesis must vanish, giving

\begin{equation}
\partial_\nu U^{\nu\mu\rho}\;=\;\frac{1}{2}\partial_\nu\left(\mathcal{S}^{\nu\mu\rho}+\mathcal{S}^{\rho\nu\mu}-\mathcal{S}^{\mu\nu\rho}\right)\;,
\end{equation}\\
where we used the conservation law $\mathcal{T}^{\mu\rho}-\mathcal{T}^{\rho\mu}=-\partial_\nu\mathcal{S}^{\nu\mu\rho}$ from Noether's first theorem, as in (\ref{TotalAngMomCons}). Integrating by parts, one last time, the last equation in (\ref{Noether2nd}) (and, thus, taking into account another ambiguity), we get the conservation law

\begin{equation}
\partial_\mu\left(\mathcal{T}^{\mu\rho}+\partial_\nu\mathcal{S}^{\mu\nu\rho}+\partial_\nu U^{\nu\mu\rho}+\partial_\nu Q^{\nu\mu\rho}\right)\;=\;0\;,
\end{equation}\\
Noticing that $\partial_\mu\partial_\nu U^{\nu\mu\rho}=0$ and requiring that the expression in the parenthesis keeps having a vanishing antisymmetric part, the new ambiguity must be

\begin{equation}
\partial_\nu Q^{\nu\mu\rho}\;=\;-\partial_\nu\mathcal{S}^{\mu\nu\rho}-\partial_\nu\mathcal{S}^{\rho\nu\mu}\;,
\end{equation}\\
which, by relabeling indices, yields the conserved current

\begin{equation}
\partial_\mu\Theta^{\mu\nu}\;=\;\partial_\mu\left(\mathcal{T}^{\mu\nu}\,+\,\frac{1}{2}\partial_\rho\left(\mathcal{S}^{\rho\mu\nu}\,+\,\mathcal{S}^{\nu\mu\rho}\,+\,\mathcal{S}^{\mu\nu\rho}\right)\right)\;=\;0\;,\label{BelinTensor2}
\end{equation}\\
a result identical to (\ref{BelinTensor}). This is Noether's second theorem, when the equations of motion are satisfied. In a sense, this is to be expected from this second theorem, since local spacetime translations (\ref{LocalTranslations}) are really the general coordinate transformations of a curved-spacetime theory with diffeomorphism invariance (whose Hilbert energy-momentum tensor has been identified as $\mathcal{T_{\tiny H}}^{\mu\nu}=\Theta^{\mu\nu}$). Note, also, that redifining the canonical currents could be made systematic and employed to extended spacetime symmetry or gauge groups \cite{Kourkoulou:2022ajr}.

\section{Duality versus translation invariance}
\subsection{Conservation of the dual energy-momentum current}\label{AppendixEMTsusy}
At first glance, spin-orbit duality could have an impact on energy-momentum conservation. This is because, from angular momentum conservation (\ref{TotalAngMomCons}), the antisymmetric part of the energy-momentum tensor is $\mathcal{T}^{[\mu\nu]}=-\frac{1}{2}\partial_\rho\,\mathcal{S}^{\rho\mu\nu}$, which, under (\ref{CurrentDuality}), transforms as

\begin{equation}
\mathcal{T}^{[\mu\nu]}\;=\;-\frac{1}{2}\partial_\rho\,\mathcal{S}^{\rho\mu\nu}\hspace{1cm}\mapsto\hspace{1cm}\tilde{\mathcal{T}}^{[\mu\nu]}\;\equiv\;-\frac{1}{4}\epsilon^{\mu\nu\alpha\beta}\partial_\rho\,{\mathcal{L}^\rho}_{\alpha\beta}\;=\;-{(\star\mathcal{T}}^{[\mu\nu]})\;.\label{DualAntiSymEMtensor}
\end{equation}\\
So what about the transformation of the symmetric part $\mathcal{T}^{(\mu\nu)}$? At this point, we notice that the above map shows how $\mathcal{T}^{[\mu\nu]}$ transforms under the duality, because it is (fundamentally) expressed in terms of the underlying transforming objects: the angular-momentum currents. Hence, we must find the analogous expression for $\mathcal{T}^{(\mu\nu)}$. That expression we already have, though, in terms of the improved energy-momentum tensor $\Theta^{\mu\nu}$ of Section \ref{SectionEMtensor}, equation (\ref{BelinTensor}), which implies
\begin{equation}
\begin{split}
\mathcal{T}^{[\mu\nu]}\;&=\;-\frac{1}{2}\partial_\rho\,\mathcal{S}^{\rho\mu\nu}\\[10pt]
\mathcal{T}^{(\mu\nu)}\;&=\;\Theta^{\mu\nu}-\frac{1}{2}\partial_\rho\left(\mathcal{S}^{\nu\mu\rho}+\mathcal{S}^{\mu\nu\rho}\right)\;.
\end{split}
\end{equation}\\
In other words, we express $\mathcal{T}^{\mu\nu}$ in terms of $\Theta^{\mu\nu}$, the latter being the fully-inclusive current that takes spin into account. If we do not do this, i.e. if we just take $\mathcal{T}^{\mu\nu}$ irrespective of $\Theta^{\mu\nu}$, we will be looking at a transformation that exchanges spin and orbit, on an object that misses its spin part. Moving on, the antisymmetric part transforms as in (\ref{DualAntiSymEMtensor}). The symmetric part, on the other hand, maps to

\begin{equation}
\mathcal{T}^{(\mu\nu)}\hspace{0.5cm}\mapsto\hspace{0.5cm}\tilde{\mathcal{T}}^{(\mu\nu)}\;=\;\tilde{\Theta}^{\mu\nu}-\frac{1}{2}\partial_\rho\left(\epsilon^{\mu\rho\alpha\beta}{\mathcal{L}^\nu}_{\alpha\beta}+\epsilon^{\nu\rho\alpha\beta}{\mathcal{L}^\mu}_{\alpha\beta}\right)\;,
\end{equation}\\
where, by a straightforward calculation,

\begin{equation}
\partial_\mu\tilde{\mathcal{T}}^{(\mu\nu)}\;=\;\partial_\mu\tilde{\Theta}^{\mu\nu}\,+\,\partial_\mu({\star\mathcal{T}}^{[\mu\nu]})\;.
\end{equation}\\
In view of (\ref{DualAntiSymEMtensor}), this means that Poincar\`e invariance in the dual theory, $\partial_\mu\tilde{\mathcal{T}}^{\mu\nu}=0$, holds if and only if $\partial_\mu\tilde{\Theta}^{\mu\nu}=0$. Hence, all we need is the transformation $\Theta^{\mu\nu}\mapsto\tilde{\Theta}^{\mu\nu}$. Nevertheless, $\Theta^{\mu\nu}\mapsto\tilde{\Theta}^{\mu\nu}$ (or $\mathcal{T}^{(\mu\nu)}\mapsto\tilde{\mathcal{T}}^{(\mu\nu)}$, for that matter) must be a map different from the previous ones, since we are now dealing with a symmetric tensor and the map needed cannot be Hodge duality. For example, we may assume the most general index-structure-preserving transformation rule $\tilde{\Theta}^{\mu\nu}={P^{\mu\nu}}_{\alpha\beta}\Theta^{\alpha\beta}$, where $P$ is symmetric in $(\mu\leftrightarrow\nu)$ and $(\alpha\leftrightarrow\beta)$, the latter being true since $\Theta^{\mu\nu}$ (and, thus, its map $\tilde{\Theta}^{\mu\nu}$) is symmetric. In turn, this also means that this tensor is a symmetric tensor product $\Theta=\sum_i\lambda_i\,v^i\otimes v^i$, for some coefficients $\lambda_i$ and in some basis $\lbrace v^i\in\mathbb{R}^{1,3}\rbrace$. Hence, whatever the exact form of the spin-orbit duality is in this case, it acts separately on the basis vectors $\lbrace v^i\rbrace$ and, thus, implies the transformation rule

\begin{equation}
\Theta^{\mu\nu}\hspace{1cm}\mapsto\hspace{1cm}\tilde{\Theta}^{\mu\nu}\;=\;{P^\mu}_\alpha{P^\nu}_\beta\,\Theta^{\alpha\beta}\;,
\end{equation}
which has to be a homomorphism, in order to preserve group structure, and a bijection, so that the inverse map exists. In fact, as summarized in Figure \ref{MAP}, a double action the spin-orbit duality should bring us back to the initial theory\footnote{As Figure \ref{MAP} shows, this is true up to a minus sign for tensors that involve/transform into the spatial part of four-position, like e.g. the angular momenta. However, here we are dealing with a symmetric tensor of second order, so even if the transformation of its $\lbrace v^i\rbrace$ basis picked up a minus sign, that would not show in its second-order tensor product. All in all, $\Theta^{\mu\nu}$ should come back to itself under the double action of the spin-orbit duality.}, which means that $P$ should be orthogonal, i.e. $P^TP=PP^T=\mathbb{1}$. In turn, that means that $P$ is an element of the Lorentz group, i.e. $P\in$ O$(1,3)$. Hence, $\Theta^{\mu\nu}\mapsto\tilde{\Theta}^{\mu\nu}$ is a homogeneous Lorentz transformation and indeed, since $\partial_\mu\Theta^{\mu\nu}=0$ in the initial theory, it also holds that $\partial_\mu\tilde{\Theta}^{\mu\nu}=0$ in the dual picture. Therefore,

\begin{equation}
\partial_\mu\tilde{\mathcal{T}}^{\mu\nu}\;=\;0\;,
\end{equation}\\
which says that energy-momentum conservation and, thus, translation invariance are still true in the dual theory. On top of that, it is straightforward to check that

\begin{equation}
\int\dd^3x\,\tilde{\Theta}^{0\nu}\;=\;\int\dd^3x\,\tilde{\mathcal{T}}^{0\nu}\;=\;\tilde{p}^\nu\;,
\end{equation}\\
as in the initial theory, which just validates that the dual energy-momentum currents keep their defining properties.

\subsection{Invariance of the Poincar\`e algebra}\label{AppendixPoincareAlgebra}
In regard to the Poincar\`e algebra, since spin-orbit duality acts on the Poincar\`e generators as $\mathbf{M}\mapsto\widetilde{\mathbf{M}}=\star\mathbf{M}$ and $\mathbf{P}\mapsto\widetilde{\mathbf{P}}=\mathbf{P}$, the relevant extension preserves the algebraic form,

\begin{equation}
\begin{split}
[\mathbf{P},\mathbf{P}]\;=\;0\hspace{1cm}&\mapsto\hspace{1cm}[\widetilde{\mathbf{P}},\widetilde{\mathbf{P}}]\;=\;[\mathbf{P},\mathbf{P}]\;=\;0\\[10pt]
[\mathbf{M},\mathbf{P}]\;=\;\eta\mathbf{P}\,-\,\eta\mathbf{P}\hspace{1cm}&\mapsto\hspace{1cm}[\widetilde{\mathbf{M}},\mathbf{P}]\;=\;\eta\mathbf{P}\,-\,\eta\mathbf{P}\;,
\end{split}
\end{equation}\\
which indicates that $\mathfrak{iso}(1,3)$ remains invariant. As with Lorentz algebra in (\ref{DualAlgebraOriginalGensExample}), we can express the dual commutators in terms of the original generators, to get maps like

\begin{equation}
\begin{split}
[\mathbf{M}_{01},\mathbf{P}_{0}]\;=\;\mathbf{P}_{1}\hspace{1cm}&\mapsto\hspace{1cm}[\widetilde{\mathbf{M}}_{01},\mathbf{P}_{0}]\;=\;[\mathbf{M}_{23},\mathbf{P}_{0}]\;=\;\mathbf{P}_{1}\\[10pt]
[\mathbf{M}_{12},\mathbf{P}_{1}]\;=\;\mathbf{P}_{2}\hspace{1cm}&\mapsto\hspace{1cm}[\widetilde{\mathbf{M}}_{12},\mathbf{P}_{1}]\;=\;[\mathbf{M}_{30},\mathbf{P}_{1}]\;=\;\mathbf{P}_{2}\;.
\end{split}\label{DualPoincareOriginalGensExample}
\end{equation}\\
This, as was with the Lorentz subalgebra, does not reflect a failure of the $\mathfrak{iso}(1,3)$ algebra. It is just an expression of the fact that, in the dual theory, Lorentz generators have (orthonormally) shifted their basis, due to Hodge duality, and are, thus, shuffled, so that the same generators in the dual picture generate different rotations. Again, this has a nice manifold interpretation through the Poincar\`e group: this is the semidirect product SO$(1,3)\rtimes\mathbb{R}^{1,3}$, which, as a set, is the Cartesian product SO$(1,3)\times\mathbb{R}^{4}$, which, in turn, is homeomorphic to $(\mathsf{RP}^3\times\mathsf{R}^3)\times\mathsf{R}^4$. Given that, as shown in the map (\ref{LorentzManifoldDuality}), spin-orbit duality is essentially an exchange between $\mathsf{RP}^3$ and $\mathsf{R}^3$ (while translations, i.e. $\mathsf{R}^4$, remain unchanged), then in the case of the Poincar\`e group the map is
\begin{equation}
(\mathsf{RP}^3\times\mathsf{R}^3)\times\mathsf{R}^4\hspace{1cm}\mapsto\hspace{1cm}(\mathsf{R}^3\times\mathsf{RP}^3)\times\mathsf{R}^4\;,
\end{equation}\\
which are just different assignments of the translations of $\mathsf{R}^4$ to elements of $\mathsf{RP}^3\times\mathsf{R}^3$, but topologically equivalent group spaces.

\section{Quantization of the position four-vector}\label{AppA}
$(1+3)$ decomposition breaks four-position into its timelike and spacelike parts,

\begin{equation}
x^\mu\;=\;\frac{(x\cdot p)p^\mu}{p^2}\,+\,{h^\mu}_\nu(p)x^\nu\;\equiv\;y^\mu\,+\,h^\mu\;,\label{Xdecomp}
\end{equation}\\
where we identify $x^\mu$ and $p^\mu$ as conjugate variables, which define the usual Poisson algebra

\begin{equation}
\lbrace x^\mu,x^\nu\rbrace_{\mbox{\tiny PB}}\;=\;\lbrace p^\mu,p^\nu\rbrace_{\mbox{\tiny PB}}\;=\;0\;,\hspace{2cm}\lbrace x^\mu,p^\nu\rbrace_{\mbox{\tiny PB}}\;=\;\eta^{\mu\nu}\;.\label{PoissonStructure}
\end{equation}
Using this canonical algebra and the decomposition (\ref{Xdecomp}), the Poisson structure breaks down into a timelike part,

\begin{equation}
\lbrace y^\mu,p^\nu\rbrace_{\mbox{\tiny PB}}\;=\;\frac{p^\mu p^\nu}{p^2}\;,\hspace{2cm}\lbrace y^\mu,y^\nu\rbrace_{\mbox{\tiny PB}}\;=\;0\;,\label{PoissonY}
\end{equation}\\
and a spacelike part,

\begin{equation}
\lbrace h^\mu,p^\nu\rbrace_{\mbox{\tiny PB}}\;=\;\eta^{\mu\nu}\,-\,\frac{p^\mu p^\nu}{p^2}\;,\hspace{2cm}\lbrace h^\mu,h^\nu\rbrace_{\mbox{\tiny PB}}\;=\;-\frac{L^{\mu\nu}}{p^2}\;,\label{PoissonH}
\end{equation}\\
both being entangled through

\begin{equation}
\lbrace y^\mu,h^\nu\rbrace_{\mbox{\tiny PB}}\;+\;\lbrace y^\nu,h^\mu\rbrace_{\mbox{\tiny PB}}\;=\;\frac{L^{\mu\nu}}{p^2}\;.\label{PoissonYH}
\end{equation}

Canonical quantization of $x^\mu$ and $p^\mu$, as usual, lifts the Poisson structure (\ref{PoissonStructure}) to the ordinary Heisenberg algebra,

\begin{equation}
[\hat{x}^\mu,\hat{p}^\nu]\;=\;i\eta^{\mu\nu}\;,\hspace{2cm}[\hat{x}^\mu,\hat{x}^\nu]\;=\;0\;,\hspace{2cm}[\hat{p}^\mu,\hat{p}^\nu]\;=\;0\;,\label{HeisenbergAlgebra}
\end{equation}\\
where we set $\hbar=1$. Then, decomposition (\ref{Xdecomp}) implies that the position operator should break into a set of two operators,

\begin{equation}
\hat{x}^\mu\;=\;\hat{y}^\mu\,+\,\hat{h}^\mu\;,\hspace{2cm}\hat{y}^\mu\;=\;\frac{(\hat{x}\cdot\hat{p})\,\hat{p}^\nu}{p^2}\;,\hspace{2cm}\hat{h}^\mu\;=\;\hat{x}^\mu\,-\,\frac{(\hat{x}\cdot\hat{p})\,\hat{p}^\nu}{p^2}\;,
\end{equation}\\
where $p^2=-m^2$ is a Casimir element of the Poincar\`e algebra that commutes with everything else and, thus, may be treated as a c-number. As far as the definition of $\hat{y}^\mu$ (and, thus, $\hat{h}^\mu$) is concerned, there is an apparent ambiguity in the position of $\hat{x}^\mu$ among the two $\hat{p}^\mu$ operators, which means that the above expressions are a particular choice of operators. However, even when we employ the Weyl quantization scheme, the most popular technique when such ambiguities come up, and average over all possible choices,

\begin{equation}
\hat{y}^\mu\;=\;\frac{1}{4}\left(\frac{(\hat{x}\cdot\hat{p})\,\hat{p}^\nu}{p^2}\;+\;\frac{(\hat{p}\cdot\hat{x})\,\hat{p}^\nu}{p^2}\;+\;\frac{\hat{p}^\nu\,(\hat{x}\cdot\hat{p})}{p^2}\;+\;\frac{\hat{p}^\nu\,(\hat{p}\cdot\hat{x})}{p^2}\right)\;,
\end{equation}\\
the resulting algebra of commutators ends up the same. In particular, this is the algebra

\begin{equation}
[\hat{y}^\mu,\hat{p}^\nu]\;=\;i\,\frac{\hat{p}^\mu\hat{p}^\nu}{p^2}\;,\hspace{2cm}[\hat{y}^\mu,\hat{y}^\nu]\;=\;0\;,
\end{equation}\\
which is, as we would anticipate, in exact correspondence with its associated Poisson subalgebra (\ref{PoissonY}).

\section{$(1+3)$ decomposition and the Newton-Wigner theorem}\label{AppendixWigner}
In reality, as mentioned in Section \ref{SubsectionFuzzy}, the correlation of our noncommutative structure with the Newton-Wigner theorem already takes place in the initial theory and it is irrespective of the context of the spin-orbit duality. That is, it originates from the $(1+3)$ decomposition of the four-position $x^\mu=y^\mu+h^\mu$, under which, as shown in Appendix \ref{AppA}, the spacelike part defines a noncommutative hyperspace,

\begin{equation}
[\hat{h}^\mu,\hat{h}^\nu]\;=\;-i\,\frac{\hat{L}^{\mu\nu}}{p^2}\;\hspace{2cm}[\hat{h}^\mu,\hat{p}^\nu]\;=\;i\eta^{\mu\nu}\,-\,i\,\frac{\hat{p}^\mu \hat{p}^\nu}{p^2}\;.\label{HyperspaceAlgebra}
\end{equation}\\
The first commutator reflects the noncommutativity of the spacelike submanifold normal to the four-momentum. Taking into account, as shown in Appendix \ref{AppA}, that $[\hat{y}^\mu,\hat{h}^\nu]+[\hat{h}^\mu,\hat{h}^\nu]=\hat{L}^{\mu\nu}/p^2$ and $[\hat{y}^\mu,\hat{y}^\nu]=0$, this noncommutative subspace is born from a nonlinear decomposition of an, otherwise, commutative space, i.e. $[\hat{x}^\mu,\hat{x}^\nu]=[\hat{y}^\mu+\hat{h}^\mu,\hat{y}^\nu+\hat{h}^\nu]=0$. The second commutator, in the low-energy regime, implies again the velocity commutator (\ref{NWvelocity}) (this time with the correct, positive sign), together with the first commutator forming the spatial subalgebra

\begin{equation}
[\hat{h}^i,\hat{h}^j]\;=\;-i\,\frac{\hat{L}^{ij}}{p^2}\;,\hspace{2cm}[\hat{h}^i,\hat{p}^j]\;=\;i\delta^{ij}\,-\,i\,\frac{\hat{p}^i\hat{p}^j}{p^2}\;.
\end{equation}\\
This algebra, in turn, constitutes a long-lived proposal of the Generalized Uncertainty Principle \cite{Maggiore:1993rv} (i.e. by first assuming the principle, they deduced the above algebra) and of the Doubly Special Relativity \cite{Amelino-Camelia:2000cpa}. More recent research on those topics \cite{Kowalski-Glikman:2002eyl} start from this kind of algebra, though sometimes restricting to commuting position operators.\\

All in all, the spacelike position $h^\mu$ defines a hyperspace normal to four-momentum $p^\mu$, introducing a $(1+3)$ slicing in spacetime, the latter foliating into equal-time three-spaces spanned by $h^i$. Those are often called the \textit{Wigner 3-spaces}. In our context, those subspaces transform in such a way in the dual theory that an overall noncommutative spacetime, i.e. (\ref{DualAlgebra}), emerges.

\section{Duality versus supersymmetry}\label{AppendixSUSY}
Like the dual Lorentz and Poincar\`e algebras, supersymmetry seems, at first, to have similar transformation properties under the spin-orbit duality. That is, considering that the supercharges $\mathbf{Q}$ themselves are unaffected by the duality, and since $\mathbf{P}\mapsto\mathbf{P}$,then
\begin{equation}
[\mathbf{M},\mathbf{Q}]\;=\;\mathbf{S}\mathbf{Q}\hspace{1cm}\mapsto\hspace{1cm}[\tilde{\mathbf{M}},\mathbf{Q}]\;=\;\mathbf{\tilde{S}}\mathbf{Q}\;,
\end{equation}
where $\mathbf{S}\mapsto\tilde{\mathbf{S}}=\star\mathbf{L}$. As was the case with the previous algebras, this map seems to preserve the form of the superalgebra. However, now, there is an important difference: this algebra separates between the $\mathbf{S}$ and $\mathbf{L}$ representations of the (total-angular-momentum) Lorentz generators $\mathbf{M}=\mathbf{S}+\mathbf{L}$. So when we express the dual superalgebra, as did with the Lorentz and Poincar\`e algebras, in terms of the original generators, we get
\begin{equation}
[\tilde{\mathbf{M}},\mathbf{Q}]\;=\;\mathbf{\tilde{S}}\mathbf{Q}\hspace{1cm}\Leftrightarrow\hspace{1cm}\star[\mathbf{M},\mathbf{Q}]\;=\;\star\mathbf{L}\mathbf{Q}\hspace{1cm}\Rightarrow\hspace{1cm}[\mathbf{M},\mathbf{Q}]\;=\;\mathbf{L}\mathbf{Q}\;.\label{DualSusyOriginalGens}
\end{equation}
In the cases of the Lorentz and Poincar\`e algebras, the analogous expressions (\ref{DualAlgebraOriginalGensExample}) and (\ref{DualPoincareOriginalGensExample}) have well-defined meanings as being the same algebras (with those of the initial theory) but in a shifted basis. On the contrary, the dual superalgebra in (\ref{DualSusyOriginalGens}) has no such properties: $\mathbf{S}$ has given its place to $\mathbf{L}$, which it is certainly not just a shift of basis. It is not clear to us whether this is an actual problem, as far as invariance of an algebra under this duality is concerned. That is, since the latter exchanges the roles between spin and orbit degrees of freedom, dual expressions like $[\tilde{\mathbf{M}},\mathbf{Q}]\;=\;\mathbf{\tilde{S}}\mathbf{Q}$ should have the same algebraic meaning with $[\mathbf{M},\mathbf{Q}]\;=\;\mathbf{S}\mathbf{Q}$ of the initial theory. In any case, assuming that it \textit{is} a problem, we may then consider that supercharges transform too under the duality,
\begin{equation}
\mathbf{Q}\hspace{1cm}\mapsto\hspace{1cm}\tilde{\mathbf{Q}}\,:\hspace{0.9cm}\mathbf{L}\tilde{\mathbf{Q}}\;=\;\star\mathbf{S}\tilde{\mathbf{Q}}\;,
\end{equation}
a map which keeps the superalgebra invariant, with the used-to change of basis. (We could also pick the map such that $\mathbf{L}\tilde{\mathbf{Q}}\;=\;\mathbf{S}\tilde{\mathbf{Q}}$ which leaves the basis alone, but we want to keep an exact analogy with the rest of the dual Poincar\`e algebra that does have a  shifted basis.) This is a statement on generators analogous to what we discussed below (\ref{SelfDualOperator}), which may be interpreted as the fact that supersymmetry in the dual theory is preserved for self-dual angular momentum. Whereas, this is an automatically satisfied condition for the dual space.\\

The strongest hint, though, that supersymmetry and the spin-orbit duality work well together, comes from superspace. In this realization, supercharges $\mathbf{Q}_\alpha=\partial_\alpha+(\sigma^\mu)_{\alpha\dot{\beta}}\theta^{\dot{\beta}}\mathbf{P}_\mu$ should stay the same, since $\mathbf{P}^\mu\mapsto\mathbf{P}^\mu$ under the duality. Then by considering Noether's theorem and deriving the corresponding supercurrent \cite{Ferrara:1974pz} we find that the latter is, as expected, dependent on the energy momentum tensor (the canonical or the improved one, whatever). In that respect, conservation of the supercurrent is not affected, since, as was illustrated, the energy-momentum conservation law is preserved under the duality. However, we also find that the supercurrent involves terms of the form $\gamma^\mu\gamma^\nu\partial_\nu$, where the gamma-matrix product is associated with both the Lorentz and Dirac spacetime algebras, as ${(\mathbf{S}^{\mu\nu})}_{\alpha\beta}=-\frac{i}{2}[\gamma^\mu,\gamma^\nu]_{\alpha\beta}$ and $\lbrace\gamma^\mu,\gamma^\nu\rbrace=2\eta^{\mu\nu}$. This yields that the product is actually a projection

\begin{equation}
\gamma^\mu\gamma^\nu\;=\;i\,\mathbf{S}^{\mu\nu}\,+\,\eta^{\mu\nu}\mathbf{\mathbb{1}}\;,
\end{equation}\\
which is obviously sensitive to the spin-orbit duality. However, in momentum space, $\gamma^\mu\gamma^\nu\partial_\nu$ reveals the spin condition $\mathbf{S}^{\mu\nu}\mathbf{P}_\nu=0$ (where the action of generators on fields is understood) and implies that only the metric part of the projection survives. Hence, all facts considered, supercurrent conservation seems to be invariant under the duality.


\end{document}